\documentclass{amsart}
\usepackage{graphicx}
\newtheorem{thm}{Theorem}[section]
\newtheorem{cor}[thm]{Corollary}
\newtheorem{lem}[thm]{Lemma}
\newtheorem{prop}[thm]{Proposition}
\theoremstyle{definition}
\newtheorem{defn}[thm]{Definition}
\theoremstyle{remark}
\newtheorem{rem}[thm]{Remark}
\numberwithin{equation}{section}
\newtheorem{exmp}[thm]{Example}
\numberwithin{equation}{section}
\newtheorem{cntxmp}[thm]{Counterexample}
\numberwithin{equation}{section}
\theoremstyle{cond}
\newtheorem{cond}[thm]{Condition}
 \numberwithin{equation}{section}
\newenvironment{prf}{ \noindent{\bf Proof}}{\\ \hspace*{\fill}$\Box$ \par  }

\newcommand{\eqdef}{\stackrel{\scriptstyle \triangle}{=}}

\newcommand{\convGH}{\raisebox{-0.5em}{$\stackrel{\longrightarrow}{\scriptstyle GH}$}}
\begin{document}

\title[Surface Triangulation -- the Metric Approach]{Surface Triangulation -- the Metric Approach }%
\author{Emil Saucan}%
\address{Department of Mathematics,Technion \& Department of Software Engineering, Ort Braude College, Karmiel}%
\email{semil@tx.technion.ac.il}%

\thanks{We would like to thank Prof. Gershon Elber of The Computer Science Department, The Technion, Haifa, who motivated and sustained this project}%

\date{24.01.2004.}%
\begin{abstract}
 We embark in a program of studying  the problem of better approximating surfaces by triangulations(triangular
meshes) by considering the approximating triangulations as finite
 metric spaces and the target smooth surface as their Haussdorff-Gromov
limit. This allows us to define in a more natural way the relevant elements, constants and invariants s.a.
principal directions and principal values, Gaussian and Mean curvature, etc. By a "natural way" we mean an
intrinsic, discrete, metric definitions as opposed to approximating or paraphrasing the differentiable notions. In
this way we hope to circumvent computational errors and, indeed, conceptual ones, that are often inherent to the
classical, "numerical" approach. In this first study we consider the problem of determining the Gaussian curvature
of a polyhedral surface, by using the {\em embedding curvature} in the sense of Wald (and Menger). We present two
modalities of employing these definitions for the computation of Gaussian curvature.
\end{abstract}
\maketitle
\section{Introduction}

The paramount importance of triangulations of surfaces and their ubiquity in various implementations (s.a. in
numerous algorithms applied in robot (and computer) vision, computer graphics and geometric modelling,
with a wide range of applications from industrial ones, to biomedical engineering to cartography and astrography -- to number just a few) has hardly to be underlined here. 
In consequence, determining the intrinsic proprieties of the surfaces under study, and especially computing their
Gaussian curvature is essential. However Gaussian curvature is a notion that is defined for smooth surfaces only,
and usually attacked with differential tools, tools that -- however ingenious and learned --  can hardly represent
good approximations for curvature of $PL$-surfaces, since they are usually just discretizations of formulas
developed in the smooth (i.e. of class at least $\mathcal{C}^2$) case.\footnote{\, However one can find very
scientifically sound discrete versions of Surface Curvature can be found, for instance, in \cite{ba2}, \cite{bcm},
\cite{c-sm}\,.}
\\ Moreover, since considering triangulations, one is faced with finite graphs, or, in many cases (when given just the vertices of the triangulation) only with
finite --thus discrete -- metric spaces. Therefore, the following natural questions arise: (A) Is one fully
justified in employing discrete metric spaces when evaluating numerical invariants of continuous surfaces? and (B)
Can one find discrete, metric equivalents of the differentiable notions, notions that are intrinsically more apt
to describe the properties of the finite spaces under investigations? One is further motivated to ask the
questions above, since the metric method we propose to employ have already successfully been used in the such
diverse fields as Geometric Group Theory, Geometric Topology and Hyperbolic Manifolds, and Geometric Measure
Theory. Their relevance to Computer Graphics in particular and Applied Mathematics in general is made even more
poignant by the study of {\it Clouds of Points} (see \cite{lwzl}, \cite{md}) and also in applications in Chemistry
(see \cite{t}).
\\ We show that the answer to both this questions is affirmative, and we focus our investigations mainly on the study
of metric equivalents of the Gauss curvature. Their role is not restricted to that of being yet another discrete
version of Gaussian Curvature, but permits us to attach a meaningful notion of curvature to points where the
surface fails to be smooth, such as {\em cone points} and {\em critical lines}. Thus we can employ curvature in
reconstructin not only smooth surface, but also surfaces with "folds", "ridges" and "facets".
\\ This exposition is organized as follows: in Section 2 we concentrate our efforts on the theoretical level and study the Lipschitz and Gromov-Hausdorff distances
between metric spaces, and show that approximating smooth surfaces by nets and triangulations is not only
permissible, but is, in a way, the natural thing to do, in particular we show that every compact surface is the
Gromov-Hausdorff limit of a sequence of finite graphs.\footnote{\, For the relevance of these notions in the study
of classical curvatures convergence, see \cite{cms}, \cite{f}\,.} In Section 3 we introduce the best candidate for
a metric (discrete) version of the classical Gauss curvature of smooth surfaces, that is the Embedding, or Wald
curvature. We study its proprieties and investigate the relationship between the Wald and the Gauss curvatures,
and show that for smooth surfaces they coincide, so that the Wald curvature represents a legitimate discrete
candidate for approximating the Gaussian curvature of triangulated surfaces. Section 4 is dedicated to developing
formulas that allow the computation of Wald curvature: first the precise ones, based upon the Cayley-Menger
determinants, and then we develop (after Robinson) elementary formulas that approximate well the Embedding
curvature. We conclude with three Appendices. In the first Appendix we present three metric analogues for the
curvature of curves, namely the Menger, Alt and Haantjes curvatures and study their mutual relationship.
Furthermore we show how to relate to these notions as metric analogues of sectional curvature and how to employ
them in the evaluation of Gauss curvature of triangulated surfaces. Next we present yet another metric analogue of
surfaces curvature, based, in this case, upon a the modern triangle comparison method, namely the Rinow curvature.
We investigate its proprieties and show (following Kirk (\cite{k})) that in the case under investigation the Rinow
and Wald curvatures coincide (and therefore Rinow curvature also identifies to the Gauss curvature). The third and
last Appendix is dedicated to the development of determinant formula for the radius of the circumscribed sphere
around a tetrahedron, with a view towards applications.

\section{The Haussdorff-Gromov limits}
\subsection{Lipschitz Distance}
\vspace*{0.2cm}This definition is based upon a very simple\footnote{\,That is to say: very intuitive, i.e. based
upon physical measurements.} idea: it measures the {\em relative} difference between metrics, more precisely it
evaluates their ratio; i.e.:
\\ The metric spaces $(X,d_X)$, $(Y,d_Y)$ are {\em close} iff  $\exists \; f:X \stackrel{\sim}{\rightarrow} Y$ s.t. $\frac{d_Y(fx,fy)}{d_X(x,y)} \approx \nolinebreak[4] 1,\, \\ \forall \ x,y \in
X$.\footnote{\,Here and in the sequel "$fx$" etc. ... stands as a short-hand version of "$f(x)$".}
\\ Technically, we give the following:
\begin{defn} The map $f:(X,d_X) \rightarrow (Y,d_Y)$ is {\em bi-Lipschitz} iff $\exists \, c,C > 0$ s.t.:
\[c\cdot d_X(x,y) \; \leq \; d_Y(fx,fy) \; \leq C \; \cdot d_X(x,y)\,.\]
\end{defn}

\begin{rem}
The same definition applies for two different metrics $d_1, d_2$ on the same space $X$.
\end{rem}
\begin{defn}
Given a Lipschitz map $f: X \rightarrow Y$, we define the {\it dilatation} of f by:
\[dil\,f = \sup_{x\neq y \in X}\frac{d_Y(fx,fy)}{d_X(x,y)} .\]
\end{defn}
\begin{rem}
The dilatation represents the {\em minimal} Lipschitz constant of maps between $X$ and $Y$.
\end{rem}
\begin{rem}
If $f$ is not Lipschitz, then $dil\,f \stackrel{\small \triangle}{=} \infty $.
\end{rem}
\begin{rem}
\begin{enumerate}
\item $f$ Lipschitz $\Rightarrow$ $f$ continuous.
 \item $f$ bi-Lipschitz  $\Rightarrow$ $f$ homeo. on its image.
\end{enumerate}
\end{rem}
\begin{rem}

We have the following results:
\begin{prop}
 Let $f,g: X \rightarrow Y$ be Lipschitz maps. Then:
\\\hspace*{3cm}(a) $g \circ f$ is Lipschitz
\\ \hspace*{2.5cm}and
\\\hspace*{3cm}(b) $dil\,(g \circ f) \leq dil\,f \cdot dil\, g$
\end{prop}
\begin{prop} The set
\(\{f:(X,d) \rightarrow \mathbb{R}\,|\, f  Lipschitz\}\) is a vector space.
\end{prop}
\end{rem}
Now we can return to our main interest and define the following notion:
\begin{defn}
 Let $(X,d_X)$, $(Y,d_Y)$  be metric spaces. Then the {\em Lipschitz distance} between $(X,d_X)$ and $(Y,d_Y)$ is
 defined as:
 \[d_L(X,Y) = \inf_{\scriptscriptstyle \stackrel{f:X \stackrel{\sim}{\rightarrow} Y}{\tiny f\; bi-Lip.}}\log{\max{(dil\,f, dil\,f^{-1})}}\]
\end{defn}
\begin{rem}
If $\neq$ $f$ bi-Lipschitz  between $X$ and $Y$, then -- remembering Remark 2.2 -- we put $d_L(X,Y)
\stackrel{\small \triangle}{=} \infty$ (i.e. $d_L$ is {\bf not} suited for pairs of spaces that are {\bf not}
bi-Lipschitz equivalent.)
\end{rem}
having defined the distance between two metric spaces we now can define the {\em convergence} in this metric in
the following natural way:
\begin{defn} The sequence of metric spaces $\{(X_n,d_n)\}$ convergence to the metric space $\{(X,d)\}$ iff
\[\lim_{n}d_L(X_n,X) = 0\]
(In this case we write: $(X_n,d_n) \raisebox{-0.2cm}{$\stackrel{\longrightarrow}{\scriptstyle L}$} 0$).
\end{defn}
\begin{exmp} Let $S_t$ be a family of regular surfaces, $S_t = f_t(U)$; where $U$ is an open set, $U = int\,U \subseteq \mathbb{R}$; such that the family $\{f_t\}$ of
parametrizations is smooth (i.e. $F:U \times \mathbb{R} \rightarrow \mathbb{R}^3 \in \mathcal{C}^1$; where
$F((u,v),t) \stackrel{\tiny not}{=} f_t(u,v)$). Then $d_n(S_t,S_0)
\raisebox{-0.15cm}{$\stackrel{\longrightarrow}{\scriptstyle t \rightarrow 0 }$}\, 0$.
\begin{rem}
If $F$ is not smooth (only continuous) then we do not necessarely have that $S_t
\raisebox{-0.15cm}{$\stackrel{\longrightarrow}{\scriptstyle t \rightarrow 0 }$} S_0$\,.
\end{rem}
\end{exmp}
We have the following significant theorem:
\begin{thm} The $d_L$ satisfies the following conditions:
\\ (a) $d_L \geq 0$;
\\ (b) $d_L$ is symmetric;
\\ (c) $d_L$ satisfies the triangle inequality;
\\ Moreover, if $X,Y$ are compact, then:
\\ (d) $d_L(X,Y) = 0 \Leftrightarrow X \cong Y$ (i.e. $X$ is {\em isometric} to $Y$);
\\that is
\\ \\ {\bf \large $d_L$ is a metric on the space of isometry classes of compact metric spaces}
\end{thm}
\begin{rem}
Let us recall the following
\begin{defn}
\((X_n,d_n) \raisebox{-0.15cm}{$\stackrel{\longrightarrow}{\scriptstyle u}$} (X,d) \stackrel{\small
\triangle}{=}\; d_u \raisebox{-0.15cm}{$\stackrel{\longrightarrow}{\scriptstyle u}$}\; d\) {\em as a real
function}; i.e.
\[\sup_{x,y \in X}|d_n(x,y) - d(x,y)| \raisebox{-0.15cm}{$\stackrel{\longrightarrow}{\scriptstyle u}$} 0\]
 (where "$u$" denotes uniform convergence.)
\end{defn}
Then $X_n \raisebox{-0.15cm}{$\stackrel{\longrightarrow}{\scriptstyle u}$} X \; \Rightarrow \; X_n
\raisebox{-0.15cm}{$\stackrel{\longrightarrow}{\scriptstyle L}$} X $ but $X_n
\raisebox{-0.15cm}{$\stackrel{\longrightarrow}{\scriptstyle L}$} X \Rightarrow \hspace{-0.4cm}{/} \; \; X_n
\raisebox{-0.15cm}{$\stackrel{\longrightarrow}{\scriptstyle u}$} X$. However, for {\em finite} spaces indeed $X_n
\raisebox{-0.15cm}{$\stackrel{\longrightarrow}{\scriptstyle u}$} X \; \Leftrightarrow \; X_n
\raisebox{-0.15cm}{$\stackrel{\longrightarrow}{\scriptstyle L}$} X $.
\end{rem}
\vspace*{0.2cm}
\subsection{Gromov-Hausdorff distance}
This is also a distance between compact metric spaces ((distinguished) up to isometry!). However it gives a {\em
weaker} topology (In particular: it is {\em always finite} (even for pairs of non-homeomorphic
spaces.)\,)\footnote{The relationship between the Lipschitz and the Hausdorff distances is akin to that between
the $\mathcal{C}^0$ and $\mathcal{C}^1$ norms in Functional Spaces.}
\\ We start by first introducing the classical
\subsubsection{Hausdorff distance}
\begin{defn}
Let $A,B \subseteq (X,d)$. We define the {\em Hausdorff distance} between $A$ and $B$ as:
\[d_H(A,B) = \inf\{r > 0\,|\, A \subset U_r(B),\, B \subset U_r(A)\}\]
(see Fig.\,1);
\begin{figure}[h]
\begin{center}
\includegraphics[scale=0.25]{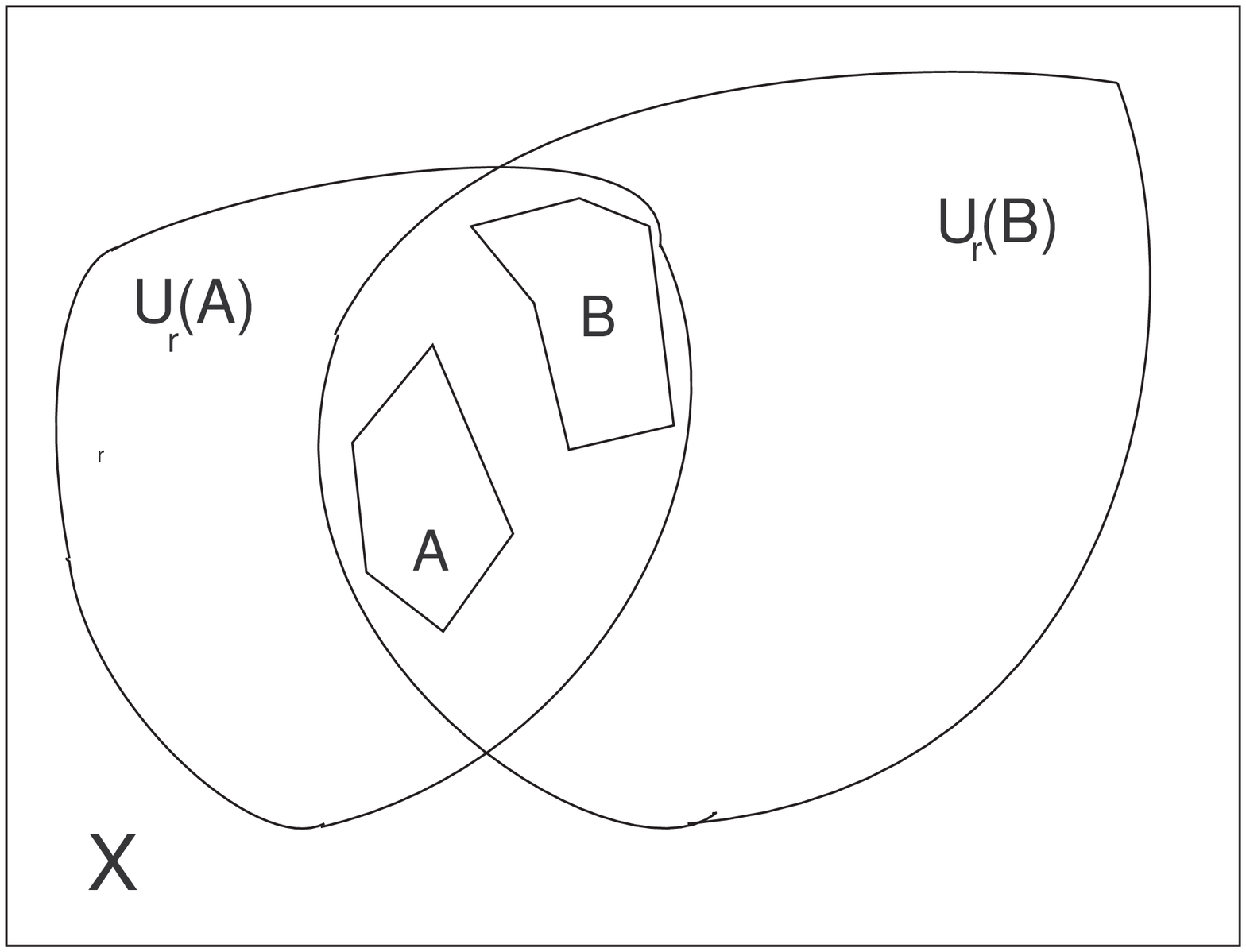}
\end{center}
\caption{ }
\end{figure}
where $U_r(A)$ is the $r$-neighborhood of A, $U_r(A) \eqdef \bigcup_{a \in A}B_r(a)$; (here, as usual: $B_r(a) =
\{x \in X \,|\, d(a,x) < r\}$.)

 Another (equivalent) way of defining the Hausdorff distance is as follows:
\[d_H(A,B) = \max\{\sup_{a\in A}d(a,B),\, \sup_{b\in B}d(b,A) \}\,.\]
(see Fig.\,2)
\begin{figure}[b]
\begin{center}
\includegraphics[scale=0.3]{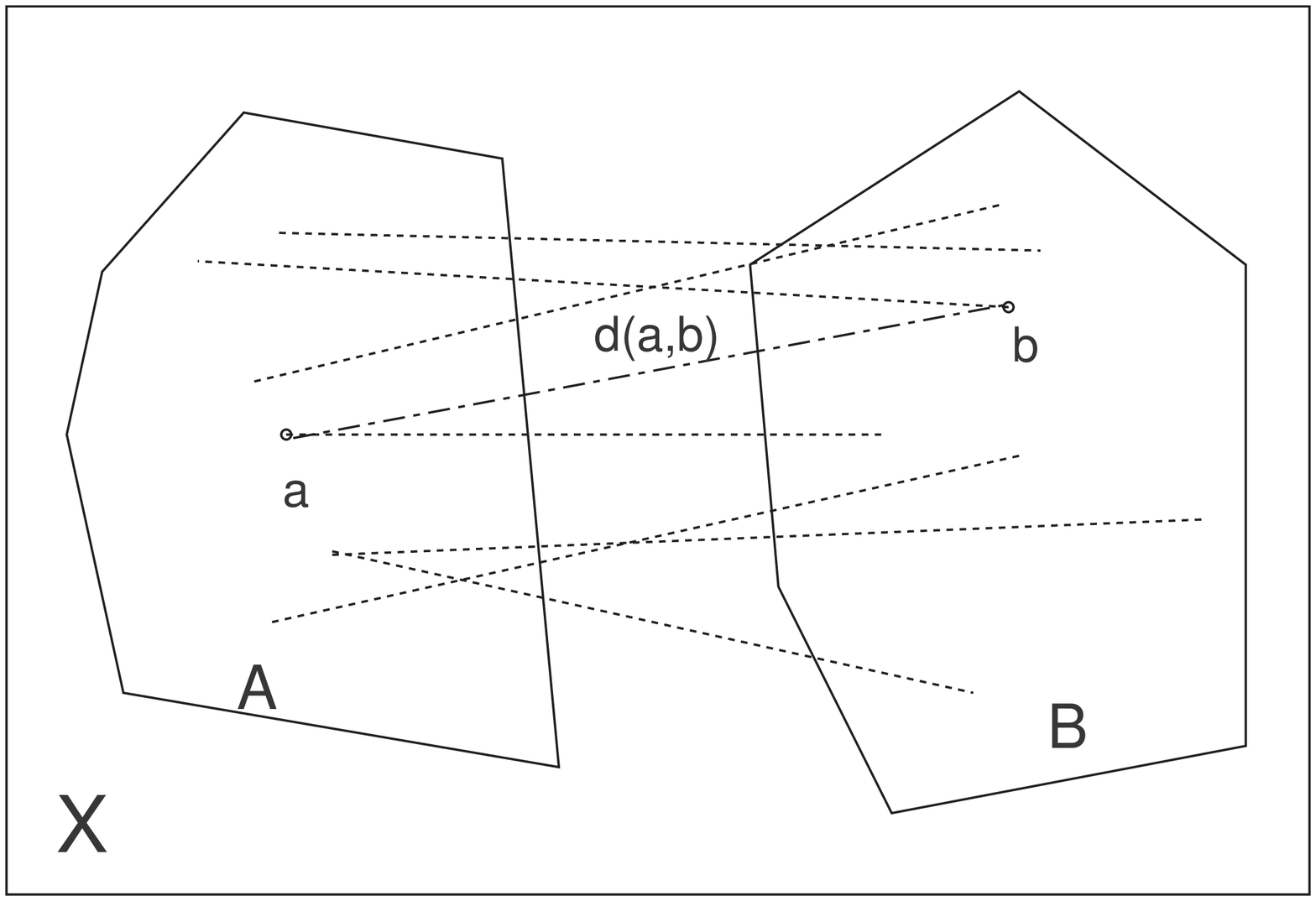}
\end{center}
\caption{ }
\end{figure}
\end{defn}
We have the following
\begin{prop}
Let $(X,d)$ be a metric space. Then:
\\ (a) $d_H$ is a semi-metric (on $2^X$). (i.e. $A = B \Rightarrow d_H{(A,B)} = 0$.)
\\ (b) $d_H(A,\bar{A}) = 0, \forall A \subseteq X$.
\\ (c) $\big(A  = \bar{A} \;and\; B = \bar{B}\big) \Rightarrow \big(d_H(A,B) = 0 \Leftrightarrow A = B\big)$.
\\ {\bf i.e. $d_H$ is a \underline{metric} on the set of closed subsets of X.}
\end{prop}
{\it  Notation} We put: $\mathcal{M}(X) = \big(\{A \subseteq X \,|\, A = \bar{A}\}, d_H\big) = 2^X \slash d_H$\,.
\begin{rem}
\begin{enumerate}
\item if $X$ is compact and if $\{A_n\}_{n \geq 1} \subseteq X$ is a sequence of compact subsets of $X$, then:
\begin{enumerate}
\item $A_{n+1} \subseteq A_n  \Rightarrow A_n \raisebox{-0.15cm}{$\stackrel{\longrightarrow}{\scriptstyle H}$} \bigcap_{n \geq
1}A_n$\,.\\
\item $A_{n} \subseteq A_{n+1}  \Rightarrow A_{n} \raisebox{-0.15cm}{$\stackrel{\longrightarrow}{\scriptstyle H}$} \bigcup_{n \geq
1}A_n$\,.\\
\end{enumerate}
\item For general subsets $A_{n} \raisebox{-0.15cm}{$\stackrel{\longrightarrow}{\scriptstyle H}$} A \in
\mathcal{M}(X)$\,, and \\
\begin{enumerate}
\item $A = \{\lim_{n}a_n \,|\, a_n \in A_n\,; n \geq 1\}$\,.\\
\item $A =  \raisebox{-0.15cm}{$\stackrel{\longrightarrow}{\scriptstyle H}$}
\bigcap_{n \geq 1}\big(\overline{\bigcup_{m=n}^{\infty}A_m }\big)$\,.
\end{enumerate}
\item If $A_n \raisebox{-0.15cm}{$\stackrel{\longrightarrow}{\scriptstyle H}$} A$, and if the sets $A_n$ are all
convex, then $A$ is convex sets.
\end{enumerate}
\end{rem}

We have the following two important results, which we present without their respective (lengthy) proofs:
\begin{prop}
$X$ complete $\Rightarrow \; \mathcal{M}(X)$ complete\,.
\end{prop}
\begin{thm}\!{\em (Blaschke)} \;$X$ compact $\Rightarrow \; \mathcal{M}(X)$ compact\,.
\end{thm}

\subsection{The Gromov-Hausdorff Distance} We are now able to define the Gromov-Hausdorff distance using the
following basic guide-lines: we want to get the {\em maximum} distance that satisfies the following two
conditions:
\\ (a) $d_{GH}(A,B) \leq d_H(A,B), \forall A,B \subset X$ (i.e. set that are close as subsets of $X$ will still be close as abstract metric
spaces);
\\and
\\ (b) $X$ isometric to $Y$ $\Longleftrightarrow$ $d_{GH}(X,Y) = 0$.
\begin{defn} Let $X,Y$ be metric spaces. Then the {\em Gromov-Hausdorff distance} between $X$ and $Y$ is defined by:
\[d_{GH}(X,Y)  \eqdef \inf d_H^Z(f(X),g(Y))\]
where the infimum is taken over all the isometric embeddings $f:X  \hookrightarrow Z, g:Y \hookrightarrow Z$ into
some metric space Z. (See Fig.\,3).
\begin{figure}[h]
\begin{center}
\includegraphics[scale=0.25]{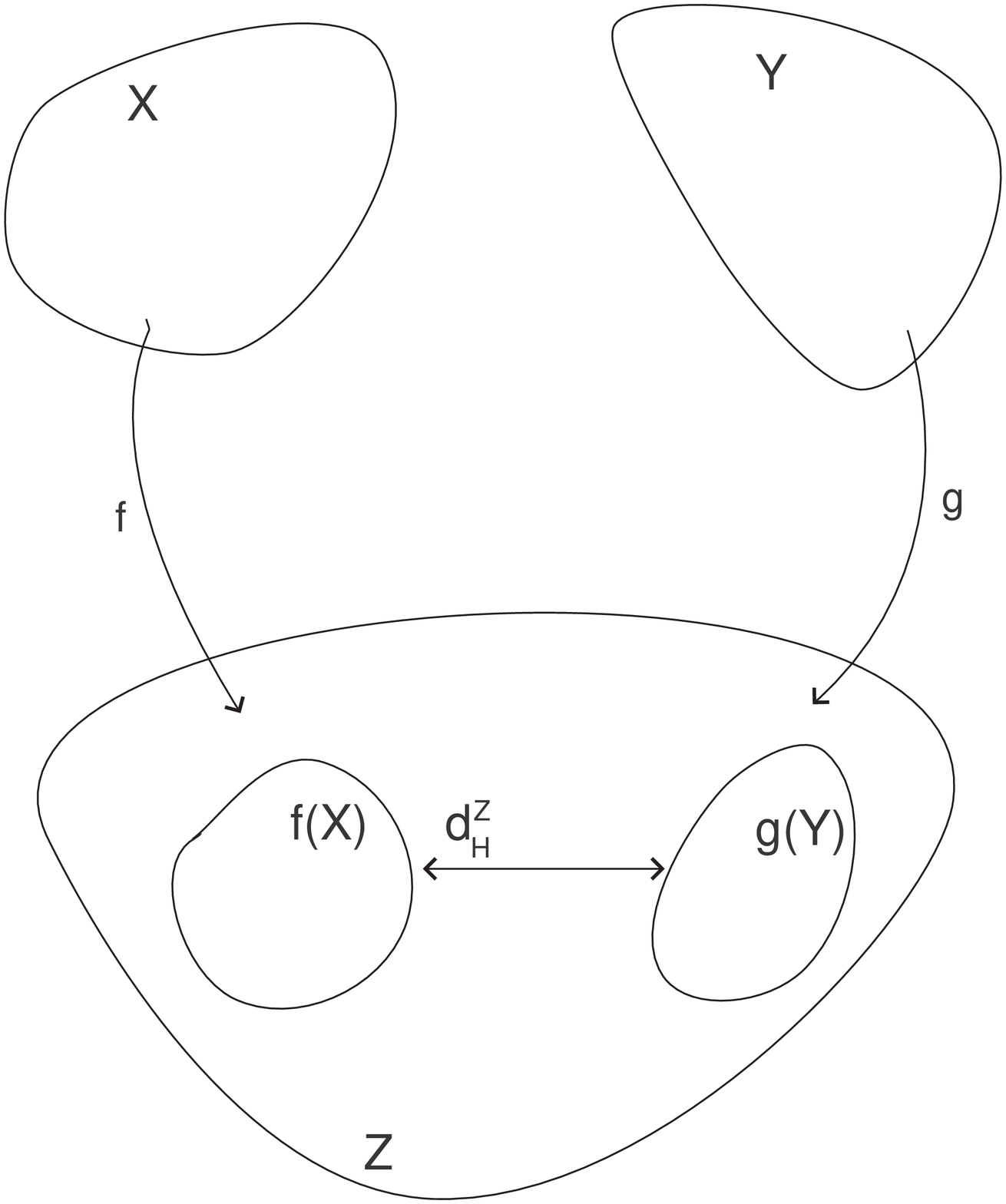}
\end{center}
\caption{ }
\end{figure}
\end{defn}
\begin{rem}
If $X = \mathbb{S}^2$, with the spherical metric, and $Z = \mathbb{R}^3$, with the Euclidian metric, then $f(X)
\neq X$\,(!)
\end{rem}
\begin{exmp}
Let $Y$ be an $\varepsilon$-net\footnote{\;{\bf Definition}
 Let $(X,d)$ be a metric space, and let $A \subset X$. $A$ is called an $\varepsilon${\em-net} iff \\ \hspace*{2.2cm}$d(x,A) \leq
 \varepsilon, \forall x \in X$.}
in $X$. Then $d_{GH}(X,Y) \leq \varepsilon$.
\\ \begin{prf}
Take $Z = X = X'\,, \; Y = Y'$.
\end{prf}
\end{exmp}
\begin{rem}
It is sufficient to consider embeddings $f$ into the disjoint union of the spaces $X$ and $Y$, $X\coprod Y$.
\end{rem}
\begin{rem}
\begin{enumerate}
\item $X,Y$ bounded $\Longrightarrow$ $d_{GH}(X,Y) \leq \infty$. \\
\item If $diamX, diamY < \infty$, then $d_{GH}(X,Y) \geq \frac{1}{2}|diamX - diamY|$\footnote{$diamX \eqdef \sup_{x,y \in X}{d(x,y)}$}.
\end{enumerate}
\end{rem}
\vspace*{0.2cm}
 However , the straightforward definition of $d_{GH}$ may be difficult to implement. Therefore we would
like to estimate (compute) $d_{GH}$ by comparing distances in $X$ vs. distances in $Y$ (as done in the cases of
uniform and Lipschitz metrics). We start by defining a {\em correspondence} between metric spaces: $X
\longleftrightarrow Y$, given by correspondences $x \leftrightarrow y$ between points $x \in X,  y \in Y$.
\begin{rem}
A correspondence is {\bf not} necessarely a function, that is to a single $x$ may correspond to several $y$-'s.
\end{rem}
We shall prove that
\[(*) \: \: \: \: d_{GH }(X,Y) < r \; \Longleftrightarrow \; \exists \;a \;correspondence X \longleftrightarrow Y\;\]
\[\hspace*{4.5cm}s.t. \; (x \leftrightarrow y, x' \leftrightarrow y') \; \Longrightarrow \; |d_X(x,x') - d_Y(y,y')| < 2r \]
Formally, we have:
\begin{defn}
Let $X,Y$ denote sets. A {\em correspondence} $X \longleftrightarrow Y$ is a subset of the Cartesian product of
$X$ and $Y$: $\mathcal{R} \subset X \times Y$ s.t.
\\(i) $\forall x \in X, \exists y \in Y, s.t. (x,y) \in \mathcal{R}$;
\\and
\\(ii) $\forall y \in Y, \exists x \in X, s.t. (x,y) \in \mathcal{R}$.
\end{defn}
\begin{exmp}
Any surjective function $f: X \rightarrow Y$ represents correspondence \\ $\mathcal{R} = \{(x,f(x))\}$.
\end{exmp}
\begin{rem}
$\mathcal{R}$ is a correspondence $\Longleftrightarrow$ $\exists\; Z$ and $\exists\; f:Z \rightarrow X\,, \exists
g:Z \rightarrow Y$; $f,g$ surjective, s.t. $\mathcal{R} = \{(f(z),g(z))\,|\, z \in Z\}$.
\end{rem}
\begin{defn}
Let $\mathcal{R}$ be a correspondence between $X$ and $Y$, where $X,Y$ are metric spaces. We Define the {\em
distortion} of $\mathcal{R}$ by:
\[dis\,\mathcal{R} = \sup\big\{\,|d_X(x,x') - d_Y(y,y')|\,\big|\, (x,y)\,,(x',y') \in \mathcal{R}\big\}\,.\]
({\small See (*)}\,.)
\end{defn}
\begin{rem}
\begin{enumerate}
\item If $\mathcal{R} = \{(x,f(x))\}$ is a correspondence induced by a surjective function $f: X \rightarrow Y$,
then $dis\,\mathcal{R} = dis\,f$, where:
\[dis\,f \eqdef \sup_{a,b \in X}|\,d_Y(fa,fb) - d_X(a,b)\,|\footnote{\,Remember that any correspondence can be expressed in this functional manner.}\,.\]
\item If $\mathcal{R} = \{(f(z),g(z))\}$, where $f:X \rightarrow Z,\; g: Y \rightarrow Z$ are surjective functions, then:
\[dis\,\mathcal{R} = \sup_{z,z' \in Z}\{\,|\,d_X(fz,fz') - d_Y(gz,gz')\,|\,\}\,.\]
\item $\mathcal{R} =  0$ iff $\mathcal{R}$ is associated to an isometry.
\end{enumerate}
\end{rem}
We bring, without proof, the following theorem:
\begin{thm}
Let $X,Y$ be metric spaces. Then:
\[d_{GH}(X,Y) = \frac{1}{2}\inf_{{\tiny \mathcal{R}} }(dis\,\mathcal{R})\,;\]
where the infimum is taken over all the correspondences $ X \stackrel{\mathcal{R}}{\longleftrightarrow} Y$.
\end{thm}
Before bringing the next result (which is very important in determining the topology ) we first introduce one more
notion:
\begin{defn}
$f:X \rightarrow Y$ is called an $\varepsilon${\em -isometry} ($\varepsilon > 0$), iff
\\(i) $dis\,f \leq \varepsilon$,
\\and
\\(ii) $f(x)$ is an $\varepsilon$-net in $Y$.
\begin{rem}
$f \; \varepsilon$-isometry $\Longrightarrow\hspace{-0.5cm}/$ \; \;$f$ continuous.
\end{rem}
\end{defn}
\begin{cor}
Let $X,Y$ be metric spaces and let $\varepsilon > 0$. Then:
\\(i) $d_{GH}(X,Y) < \varepsilon  \Longrightarrow \exists \; 2\varepsilon{-isometry} \; f:X \rightarrow Y$.
\\(ii) $\exists \; \varepsilon{-isometry} \; f:X \rightarrow Y \Longrightarrow d_{GH}(X,Y) < 2\varepsilon$.
\end{cor}
\hspace*{-0.3cm}{\bf Proof} (i) Let $ X \stackrel{\mathcal{R}}{\longleftrightarrow} Y$ s.t. $dis\,\mathcal{R} <
2\varepsilon$.
\\For any $x \in X$ and $f(x) \in Y$, choose $y = f(x)$ s.t. $(x,f(x)) \in \mathcal{R}$. Then $x \mapsto f(x)$
defines a map $f:X \rightarrow Y$. Moreover: $dil\,f \leq dil\,\mathcal{R} < \varepsilon$.
\\ We shall prove that $f(X)$ is a $2\varepsilon$-net in $Y$.
\\Indeed, let $x \in X$ and $y \in Y$ s.t. $(x,y) \in \mathcal{R}$.
Then $d(y,fx) \leq d(x,x) +  dis\,\mathcal{R} < 2r$, thence $d(y,f(X)) < 2r$.
\\ \hspace*{11cm} $\square$
\\ Let $f$ be an $2\varepsilon$-isometry. Define $\mathcal{R} \subset X \times Y, \; \mathcal{R} = \{(x,y)\,|\, d(y,fx) \leq \varepsilon\}$.
\\ Then, since $f(X)$ is an $\varepsilon$-net it follows that $\mathcal{R}$ is a correspondence.
\\ Then $\forall\, (x,y),(x',y') \in \mathcal{R}$ we have:
\[|\,d_Y(y,y') - d_X(x,x')\,| \leq |\,d(fx,fx') - d(x,x')\,| +  d(y,fx) + d(y',fx') \leq  dis\,f + \varepsilon + \varepsilon \leq 3\varepsilon \,.\]
\[\Longrightarrow dis\,\mathcal{R} \leq 3\varepsilon \;\Longrightarrow \; d_{GH}(x,y) \leq 3r/2 < 2r\,.\]
 \hspace*{11cm} $\square$ \\
 The next result is of great importance (in particular so in our context):
\begin{thm}
$d_{GH}$ is a (finite) metric on the set of isometry classes of compact metric spaces.\end{thm}
\begin{prf}
It suffices to prove that $d_{GH}(X,Y) = 0 \Longrightarrow X \stackrel{\sim}{\equiv} Y$.\footnote{We shall write:
$X \stackrel{\sim}{\equiv} Y$ if $X$ is isometric to $Y$.}
\\ Indeed, let $X,Y$ be compact spaces s.t. $d_{GH} = 0$. Then it follows from the previous Corollary (for $\varepsilon =
1/n$) that $\exists \;(f_n)_{n \geq 1}, \,f_{n} : X \rightarrow Y$ s.t. $dis\,f_n
\raisebox{-0.3em}{$\stackrel{\longrightarrow}{\scriptstyle n}$} 0$.
\\ let $S \subset X, \; \bar{S}\,, \; |S| = \aleph_0$. Using a Cantor-diagonal argument one easily shows that
$\exists \;(f_{n_k})_{k \geq 1} \subset (f_n)_{n \geq 1}$ s.t. $(f_{n_k})_{ \geq 1}$ converges in $Y, \forall x
\in S$. Without restricting the generality we may assume that this happens for $(f_n)_{n \geq 1}$  itself. Thus we
can define a function $f:X \rightarrow Y$ by putting: $f(x) = \lim_{n}f_n(x)$.
\\But $|\,d(f_nx,f_ny) - d(x,y)\,| \leq dis\,f_n  \raisebox{-0.3em}{$\stackrel{\longrightarrow}{\scriptstyle n}$} 0\; \Longrightarrow \;d(fx,fy) = \lim
d(f_nx,f_ny)$. In other words $f|\raisebox{-0.2em}{$S$}$ is an isometry. But $S = \bar{S}$, therefore this
isometry can be extended to an isometry $\tilde{f}$ from $X$ to $Y$. In a analogous manner one shows the existence
of an isometry $\tilde{\tilde{f}}:X \rightarrow Y$.
\end{prf}
\begin{rem}
$X_n\raisebox{-0.4em}{$\stackrel{\longrightarrow}{\scriptstyle L}$} X \;\Longrightarrow\;
X_n\raisebox{-0.3em}{$\stackrel{\longrightarrow}{\scriptstyle
u}$}X\;\Longrightarrow\;X_n\raisebox{-0.4em}{$\stackrel{\longrightarrow}{\scriptstyle GH}$}X $.
\end{rem}
In fact, the  following relationship exists between "$\raisebox{-0.4em}{$\stackrel{\longrightarrow}{\scriptstyle
L}$}$" and "$\raisebox{-0.4em}{$\stackrel{\longrightarrow}{\scriptstyle GH}$}$":
\begin{thm}
 $X_n \raisebox{-0.5em}{$\stackrel{\longrightarrow}{\scriptstyle GH}$} X  \;\Longleftrightarrow \; \;
\varepsilon$-nets in  $X_n  \raisebox{-0.5em}{$\stackrel{\longrightarrow}{\scriptstyle L}$} \varepsilon$-nets in
$X$.
\end{thm}
One can formulate this assertion in a more formal manner and it directly (see \cite{g+}, pg. 73). However we shall
proceed in more "delicate" manner, starting with:
\begin{defn}
Let $X,Y$ be compact metric spaces, and let $\varepsilon, \delta > 0$. $X,Y$ are called
$\varepsilon$-$\delta${\em-approximations} (of each-other) iff: $\exists$ $\{x_i\}_{i=1}^N \subset X$, $\exists$
$\{y_i\}_{i=1}^N \subset Y$ s.t.
\\ (i) $\{x_i\}_{i=1}^N $ is an $\varepsilon$-net in $X$ and $\{y_i\}_{i=1}^N $ is an $\varepsilon$-net in $Y$;
\\ (ii) $|\,d_X(x_i,x_j) - d_(y_i,y_j)\,| < \delta \; \forall\, i,j \in \{1,...,N\}$.
\\An $(\varepsilon,\varepsilon)$-approximation is called, for short: an $\varepsilon${\em-approximation}.
\end{defn}
The relationship between this last definition and the Gromov-Hausdorff distance is first revealed in
\begin{prop}
Let $X,Y$ be compact metric spaces. Then:
\begin{enumerate}
\item If $Y$ is a $(\varepsilon,\delta)$-approximation of $X$, then $d_{GH}(X,Y) < 2\varepsilon + \delta$.
\item $d_{GH}(X,Y) < \varepsilon \; \Longrightarrow \; Y$ is a $5\varepsilon$-approximation of $X$.
\end{enumerate}
\end{prop}
\begin{prf} (1) Condition (ii) of Def. 2.41. is equivalent to $dis\,\mathcal{R}_{X_0Y_0} < \delta$, where \\$X_0 = \{x_i\}_{i=1}^N\,,
\{y_i\}_{i=1}^N$. But $dis\,\mathcal{R}_{X_0Y_0} < \delta \;\Longrightarrow\; d_{GH}(X_0,Y_0) < \delta/2$. Now,
since $X_0$ and $Y_0$ are $\varepsilon$-nets in $X$, resp. $Y$, it follows that $d_{GH}(X,X_0) \leq \varepsilon,
d_{GH}(Y,Y_0) <  \varepsilon$. From here and from the $d_{GH}(X_0,Y_0) < \delta/2$ follows, by means of he
triangle inequality, that $d_{GH}(X,Y) < 2\varepsilon + \delta$.
\\ \hspace*{12.25cm} $\square$
\\ (2) By Cor. 2.37., there exists a $2\varepsilon$-isometrie $f:X \rightarrow Y$. Let $X_0 = \{x_i\}_{i=1}^N $ be
an $\varepsilon$-net, and let $y_i= f(x_i)$.
\\Then $|\,d(x_i,x_j) - d_(y_i,y_j)\,| < 2\varepsilon < 5\varepsilon$. Therefore suffice to prove that $Y_0 = \{y_i\}_{i=1}^N$
is a $5\varepsilon$-net in $Y$.
\\Indeed, if $y \in Y$, then, since $f(X)$ is an $2\varepsilon$-net in $Y$, $\exists\, x \in X$ s.t. $d(y,f(x))$.
Now, since $X_0$ is an $\varepsilon$-net in $X$, $\exists x_i \in X_0$, s.t. $d(x,x_i) \leq \varepsilon$.
\\ Therefore: $d(y,y_i) = d(y,f(x_i)) \leq d(y,f(x)) + d(f(x),f(x_i))$
\\  \hspace*{5cm}$\leq 2\varepsilon + d(x,x_i) + dis\,f$
\\  \hspace*{5cm}$\leq 2\varepsilon  + \varepsilon  + 2\varepsilon  \leq 5\varepsilon$.
\end{prf}
\begin{rem}\;
Prop. 2.42. $\Longleftrightarrow$ $(X_n\raisebox{-0.5em}{$\stackrel{\longrightarrow}{\scriptstyle GH}$}X)$
$\Leftrightarrow$ ($\forall \,\varepsilon > 0, \; X_n$ is an $\varepsilon$-approximation,
\hspace*{8,7cm}$\forall\,n$ large enough.)
\end{rem}
More precisely we have the following Proposition:
\begin{prop} Let $X, \{X_n\}_{1}^{\infty}$ compact metric spaces. Then:
\\ $X_n \convGH X$ $\Longleftrightarrow$ $\forall \varepsilon > 0,\; \exists$ a {\em finite} $\varepsilon$-net $S \subset X$
and $\exists$ a {\em finite} $\varepsilon$-net $S_n \subset X_n$, s.t. $S_n \convGH S$ and, moreover, $|S_n| =
|S|$, for large enough $n$.
\end{prop}
\begin{prf}
($\Longleftarrow$) If $S,S_n$ exist, then $\Longrightarrow \; X_n$ is an $\varepsilon$-approximation of $X$
$\stackrel{2.42.}{\Longrightarrow} \; X_n \convGH X$ \nolinebreak[4] $\forall\,n$. \\ \hspace*{6cm}
\\ \hspace*{12.3cm} $\square$
\\ \hspace*{0.9cm} ($\Longrightarrow$) Let $S$ be an {\em finite} $\varepsilon/2$-net in $X$.
\\ We  construct in $X_n$ corresponding nets $S_n$ (to be more precise, we define: $S_n = f_n(X)$, where $f_n$ is an $\varepsilon_n$-approximation, $f_n:X \rightarrow X_n, \; \varepsilon_n \rightarrow 0$.)
Then $S_n \convGH S$ and, in addition, $S_n$ is an $\varepsilon$-net in $S$ (for $n$ large enough).
\end{prf}
We make the following extremely important Remark:
\begin{rem} Let $\mathcal{M}(n,k,D)$ be an $n$-dimensional manifold, of (sectional, Ricci) curvature $\leq k$, and
s.t. $diam\,\mathcal{M} \geq D$. Then $(\mathcal{M},d_{GH})$ is compact. However, it should be noted that this
result doesn't hold for curvature $< k$. (only for $Vol(\mathcal{M}) \leq V_0)$ {\em and} injectivity radius $\geq
r_0 $.
\end{rem}
{\bf Note} With the notations of the precedent Proposition, the distances in $S_n$ converge to the distances in
$S$, as $X_n \convGH  X$, therefore {\bf The Geometric Proprieties of $S_n$ will converge to those of $S$}. Thus
{\bf we can use the Gromov-Hausdorff each and every time The  Geometric Proprieties of $X_n$ can be expressed in
term of a finite number of points, and, by passing to the limit, automatically obtain proprieties of $X$.}
\\ A typical example is that of the {\em intrinsic metric} i.e. {\em the metric induced by a length structure (i.e. path length) by a metric on a subset (of a given metric space).} (See
Fig.\,4 for the classical example of surfaces in $\mathbb{R}^3$.)
\begin{figure}[h]
\begin{center}
\includegraphics[scale=0.3]{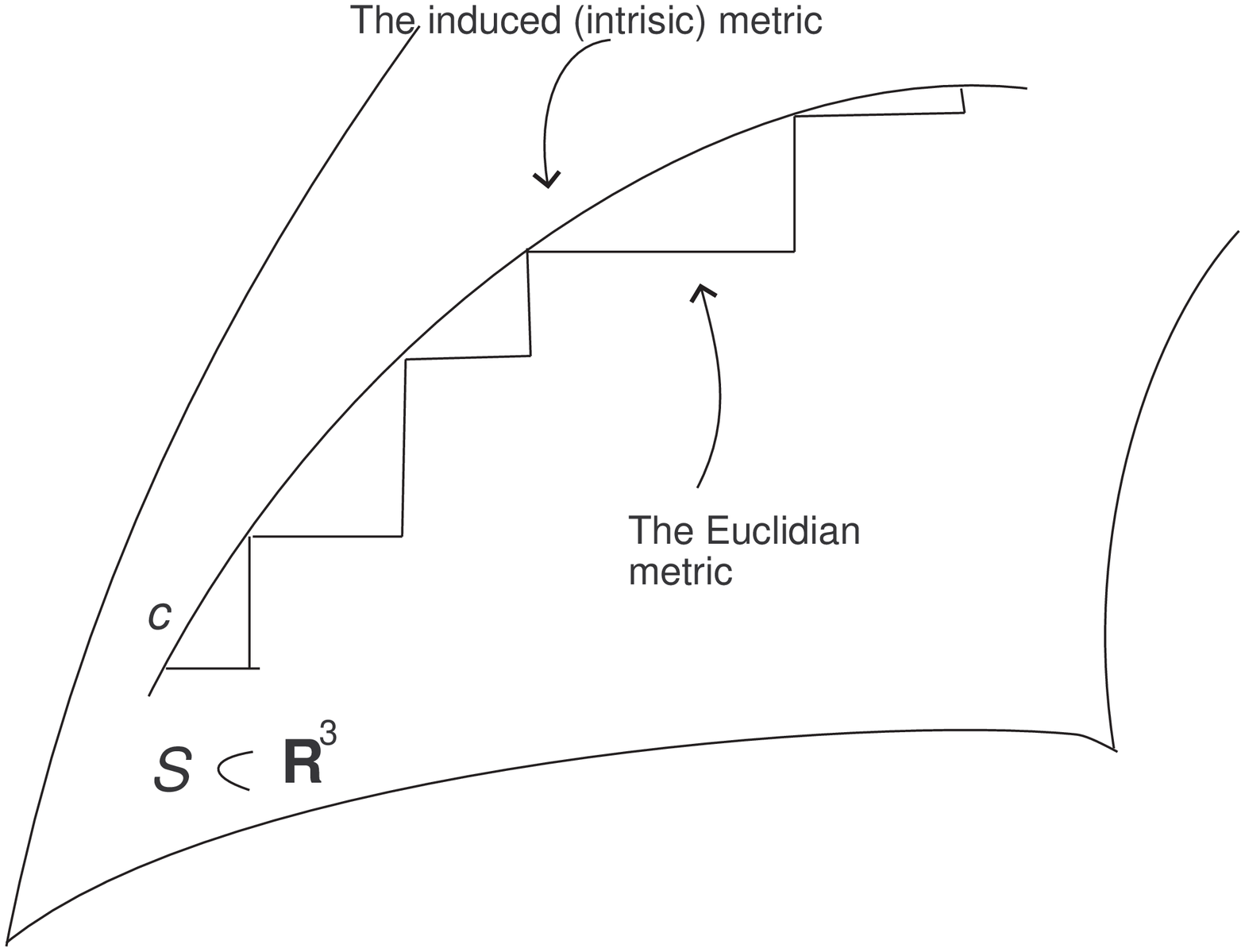}
\end{center}
\caption{ }
\end{figure}
\\ On a more formal note, we have the following characterization of intrinsic metrics:
\begin{thm} Let $(X,d)$ be a {\em complete} metric space.
\begin{enumerate}
\item If $\forall \, x,y \in X, \; \exists \frac{1}{2}xy$, then $d$ is strictly intrinsic.
\item If $\forall \, x,y \in X$ and $\forall \varepsilon > 0,\; \exists$ the $\varepsilon$-middle of $xy$, then $d$
is intrinsic.
\end{enumerate}
\end{thm}
Where we used the following definitions and notations:
\begin{defn}
\begin{enumerate}
\item Given $x,y$ points in $(X,d)$, {\em the middle} (or {\em midpoint}) of the {\em segment} $xy$ (more correctly: 'a
midpoint between "$x$" and "$y$"\,') is defined as: \[\frac{1}{2}xy = z,\; d(x,z) = d(z,y)\,.\]
\item $d$ is called {\em strictly intrinsic} iff the length structure is associated with is complete.
\item Let $d$ be an intrinsic metric. $z$ is an $\varepsilon${\em -middle} (or an $\varepsilon${\em -midpoit}) for
$xy$ iff: \\ \(|\,2d(x,z) -d(x,y)\,| \leq \varepsilon\) and \(|\,2d(y,z) -d(x,y)\,| \leq \varepsilon\).
\end{enumerate}
\end{defn}
\begin{rem} The converse of Thm. 2.46. holds in any metric space, more precisely we have:
\begin{prop} If $d$ is an intrinsic metric, then $\frac{1}{2}xy$ exists, $\forall x,y$.
\end{prop}
\end{rem}
The following Theorem shows that length spaces are closed in the {\em GH-topology}\,:
\begin{thm}
Let $\{X_n\}$ be length spaces and let $X$ be a complete metric space \hspace*{2.8cm}s.t. $X_n \convGH X$.
\\ \hspace*{2.7cm}Then $X$ is a length space.
\end{thm}
\begin{prf} We have already presented the idea of the proof: it is sufficient to show that for every $x,y$ there
exist an $\varepsilon${\em -midpoit} ($\forall \varepsilon >0$).
\\ Indeed, let $n$ be such that $d_{GH} < \frac{\varepsilon}{10}$. Then, from the a preceding result, it follows
that there exist a correspondence $X_n \raisebox{-0.1em}{$\stackrel{\scriptstyle
\mathcal{R}}{\longleftrightarrow}$} X$ s.t. $dis\,\mathcal{R} < \frac{\varepsilon}{5}$.
\\ Let $\bar{x},\bar{y} \in X_n$, $x \raisebox{-0.1em}{$\stackrel{\scriptstyle \mathcal{R}}{\leftrightarrow}$}
\bar{x}$, $y \raisebox{-0.1em}{$\stackrel{\scriptstyle \mathcal{R}}{\leftrightarrow}$} \bar{y}$. Since $X_n$ is a
length space, $\Longrightarrow \exists \bar{z} \in X_n$ s.t. $\bar{z}= \frac{\varepsilon}{5}$-midpoint of
$x_ny_n$. Consider $z \in X, \;  z \raisebox{-0.1em}{$\stackrel{\scriptstyle \mathcal{R}}{\leftrightarrow}$}
\bar{z}$. Then:
\[\big| |xz| - \frac{1}{2}|xy| \big| \leq \big| |\bar{x}\bar{z}| - \frac{1}{2}|\bar{x}\bar{y}| \big| + 2dis\,\mathcal{R} < \frac{\varepsilon}{5} + \frac{2\varepsilon}{5} < \varepsilon\,.\]
(Here we write $|xy|$ instead of $d(x,y)$, etc.)
\\ In a similar manner we show that: $\big| |yz| - \frac{1}{2}|xy| \big| < \varepsilon$; i.e. $\varepsilon${\em
-midpoit} of $xy$.
\end{prf}
The next Theorem and its Corollary are of paramount importance:
\begin{thm} Any compact length space is the GH-limit of a sequence of finite graphs.
\end{thm}
\begin{prf} Let $\varepsilon, \delta\; (\delta \ll \varepsilon)$  small enough, and let $S$ be a $\delta$-net in
$X$.
\\ Let $G = (V,E)$ be the graph with $V = S$ and $E = \{(x,y)\,|\, d(x,y) <  \varepsilon\}$. we shall prove that
$G$ is an $\varepsilon$-approximation of $X$, for $\delta$ small enough (i.e. for $\delta <
\frac{\varepsilon^2}{4}\,diam(X)$). (See Fig.\,5.)
\begin{figure}[h]
\begin{center}
\includegraphics[scale=0.2]{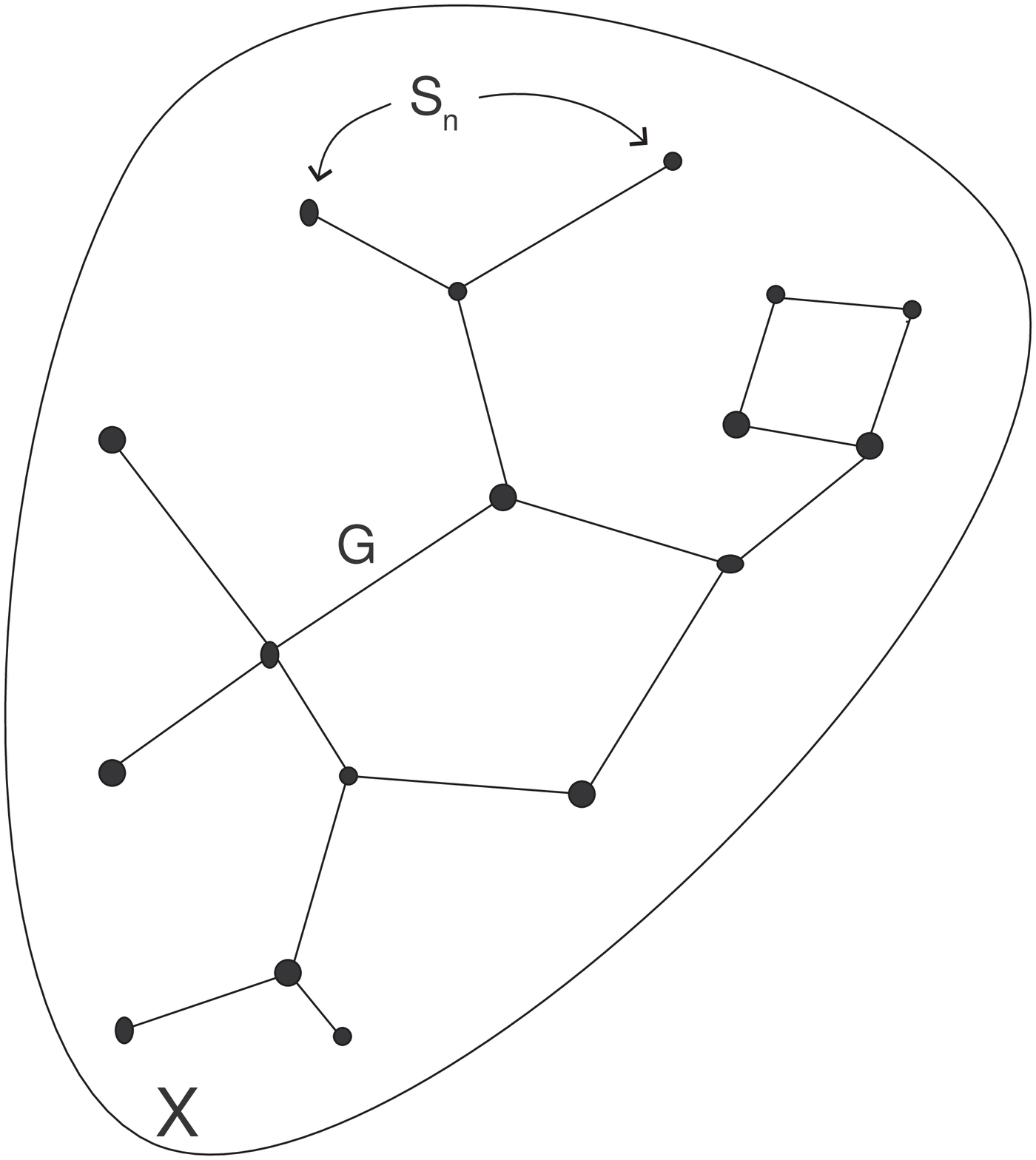}
\end{center}
\caption{ }
\end{figure}
\\ But, since $S$ is an $\varepsilon$-net both in $X$ {\bf and} in $G$, and since $d_G(x,y) \geq d_X(x,y)$, it is sufficient to prove
that: \[d_G(x,y) \leq d_X(x,y) + \varepsilon \,.\] Let $\gamma$ be the shortest path between $x$ and $y$, and let
$x_1,...,x_n \in \gamma$ s.t. $n \leq length(\gamma)/\varepsilon$ (and $|x_i,x_{i+1})|\leq \varepsilon/2)$. Since
$\forall\, x_i\, \exists y_i \in S$ s.t. $|x_i,y_i| \leq \delta$, it follows that  \(|y_iy_{i+1}| \leq
|x_ix_{i+1}| + 2\delta < \varepsilon.\) (See Fig.\,6)
\\ Therefore, (for $\delta < \varepsilon/4$) $\exists$ an edge $e \in G, e = y_iy_{i+1}$. From this we get the
following upper bound for $d_G(x,y)$:
\[d_G(x,y) \leq \Sigma_{0}^{n}|y_iy_{i+1}| \leq \Sigma_{0}^{n}|x_ix_{i+1}| + 2\delta n\]
 But $n < 2length(\gamma)/\varepsilon \leq 2diam(X)/\varepsilon$; therefore:
\[  d_G(x,y) \leq |xy| + \delta\frac{4diam(X)}{\varepsilon} < |xy| + \varepsilon\,.\]
\begin{figure}[h]
\begin{center}
\includegraphics[scale=0.3]{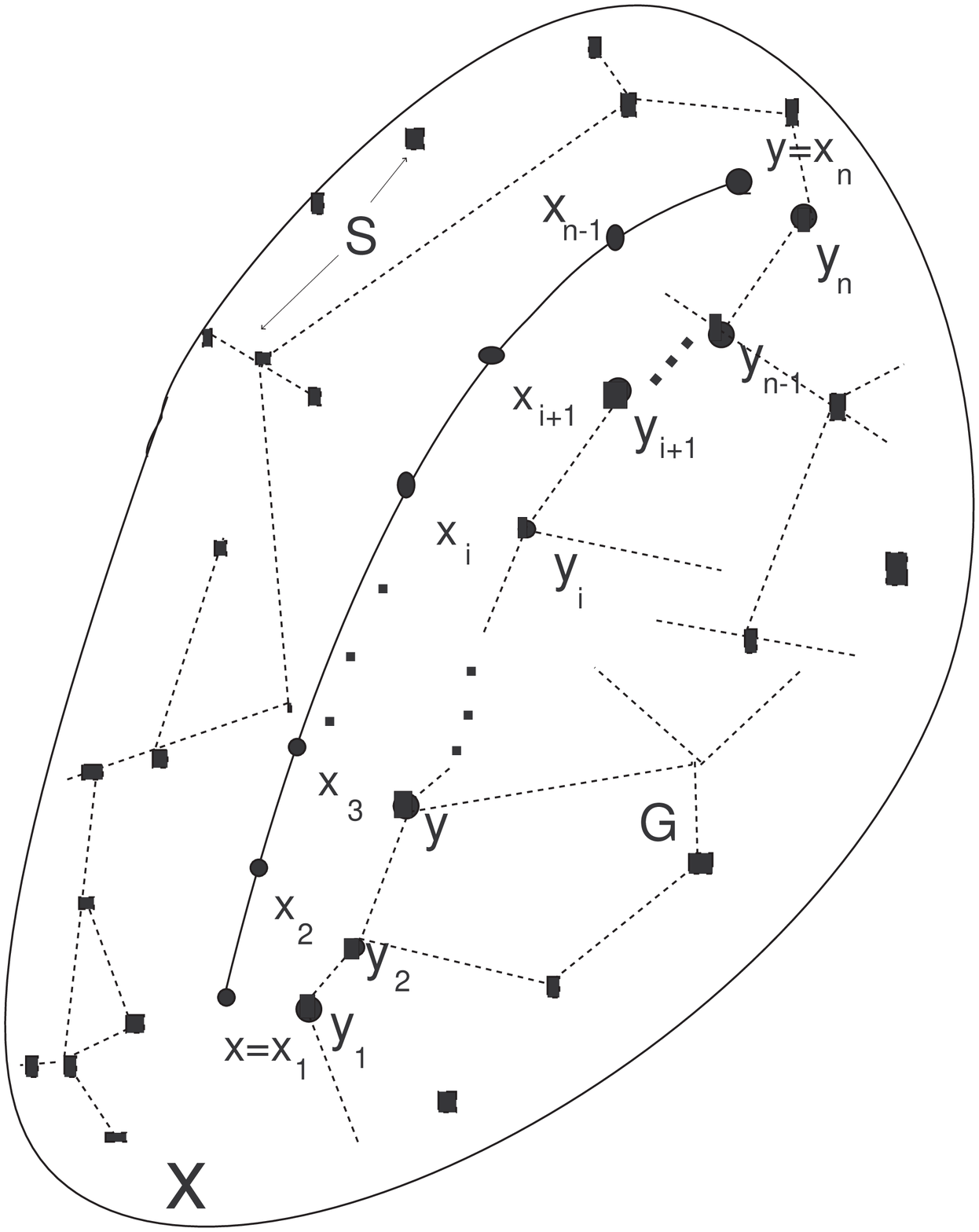}
\end{center}
\caption{ }
\end{figure}
(because $\delta < \varepsilon^2/4diam(X)$).
\\  So, for any $\varepsilon > 0, \; \exists\, G = G_{\varepsilon}$  an $\varepsilon$-approximation of $X$.
Then, $G_n  \raisebox{-0.3em}{$\stackrel{\rightarrow}{\scriptstyle \varepsilon}$} X$.
\end{prf}
\begin{cor} Let $X$ be a compact length space. Then $X$ is the Gromov-Hausdorff limit of a sequence $\{G_n\}_{n\geq1}$ of
finite graphs, isometrically embedded in $X$.
\end{cor}
\begin{rem}
\begin{enumerate}
\item If $G_n  \raisebox{-0.3em}{$\stackrel{\rightarrow}{\scriptstyle \varepsilon}$} X$, $G_n = (V_n,E_n)$. If $\exists N_0 \in \mathbb{N}$ s.t.
 \[(\star)\;\;\; |E_n| \leq N_0, \; \forall n \in \mathbb{N}\,,\]
then $X$ is a finite graph.
\item If condition $(\star)$ is replaced by:
\[(\star\star) \;\;\; |V_n| \leq N_0, \; \forall n \in \mathbb{N}\,,\]
then $X$ will still be always a graph, but not necessarily finite(!)
\end{enumerate}
\end{rem}

\section{The Embedding Curvature}
\subsection{Theoretical Setting}
This is basically a {\em comparison-curvature} (as is the more "modern" $CAT$\footnote{\,i.e.
Cartan-Alexandrov-Topogonov} approach). This is done with quadruples instead of triangles (like in the
Alexandrov-Topogonov method). It is in a sense a more natural idea, since quadruples are classically\footnote{\,as
illustrated by the time-honored principles of Projective Geometry...} the "minimal" geometric figures that allow
the differentiation between metric spaces. This allows for a much more easier and rapid development of the theory
than the triangle-based comparison. Moreover we shall show that the two Theories coincide on those metric space on
which both can be applied, i.e. metric spaces that are (a) "planar" and (b) "rich enough" i.e. contain
quadrangles, s.a. classical (PL-smooth) surfaces in $\mathbb{R}^3$.\footnote{\,In this sense CAT spaces are more
"potent": they can be employed in studying mathematical objects that not (neccessarilly) contain quadrangles, e.g.
trees, Cayley graphs, etc..}

\begin{defn}
Let $(M,d)$ be a metric space, and let $Q = \{p_1,...,p_4\} \subset M$, together with the mutual distances:
$d_{ij} = d_{ji} = d(p_i,p_j); \, 1 \leq i,j \leq 4$. The set $Q$ together with the set of distances
$\{d_{ij}\}_{1\leq i,j \leq 4}$ is called a {\em metric quadruple}.
\end{defn}

\begin{rem} One can define metric quadruples in slightly more abstract manner, without the aid of the ambient
space: a metric quadruple being a $4$ point metric space; i.e. $Q = \big(\{p_1,...,p_4\}, \{d_{ij}\}\big)$, where
the distances $d_{ij}$ verify the axioms for a metric.
\end{rem}
Before we proceed to the next definition, let us introduce the following
\\ \\ {\em Notation} $S_{\kappa}$ denotes {\em the complete, simply connected surface of constant curvature $\kappa$}, i.e. $S_{\kappa} \equiv \mathbb{R}^2$, if $\kappa = 0$; $S_{\kappa} \equiv \mathbb{S}^2_{\sqrt{\kappa}}$\,, if $\kappa > 0$; and $S_{\kappa} \equiv \mathbb{H}^2_{\sqrt{-\kappa}}$
 \,,\;if $\kappa < 0$. Here $S_{\kappa} \equiv \mathbb{S}^2_{\sqrt{\kappa}}$ denotes the Sphere of radius
 $R = 1/\sqrt{\kappa}$, and $S_{\kappa} \equiv \mathbb{H}^2_{\sqrt{-\kappa}}$ stands for the Hyperbolic Plane of
 curvature $\sqrt{-\kappa}$, as represented by the Poincare Model of the plane disk of radius $R =
 1/\sqrt{-\kappa}$ 
\begin{defn} The {\em embedding curvature} $\kappa(Q)$ of the metric quadruple $Q$ is defined  be the curvature $\kappa$ of
$S_{\kappa}$ into which $Q$ can be isometrically embedded. (See Figures 7 and 8 for embeddings of the metric
quadruple in $S_{\kappa}$ and $H_{\kappa}$, respectively.)
\end{defn}

\begin{figure}[h]
\begin{center}
\includegraphics[scale=0.3]{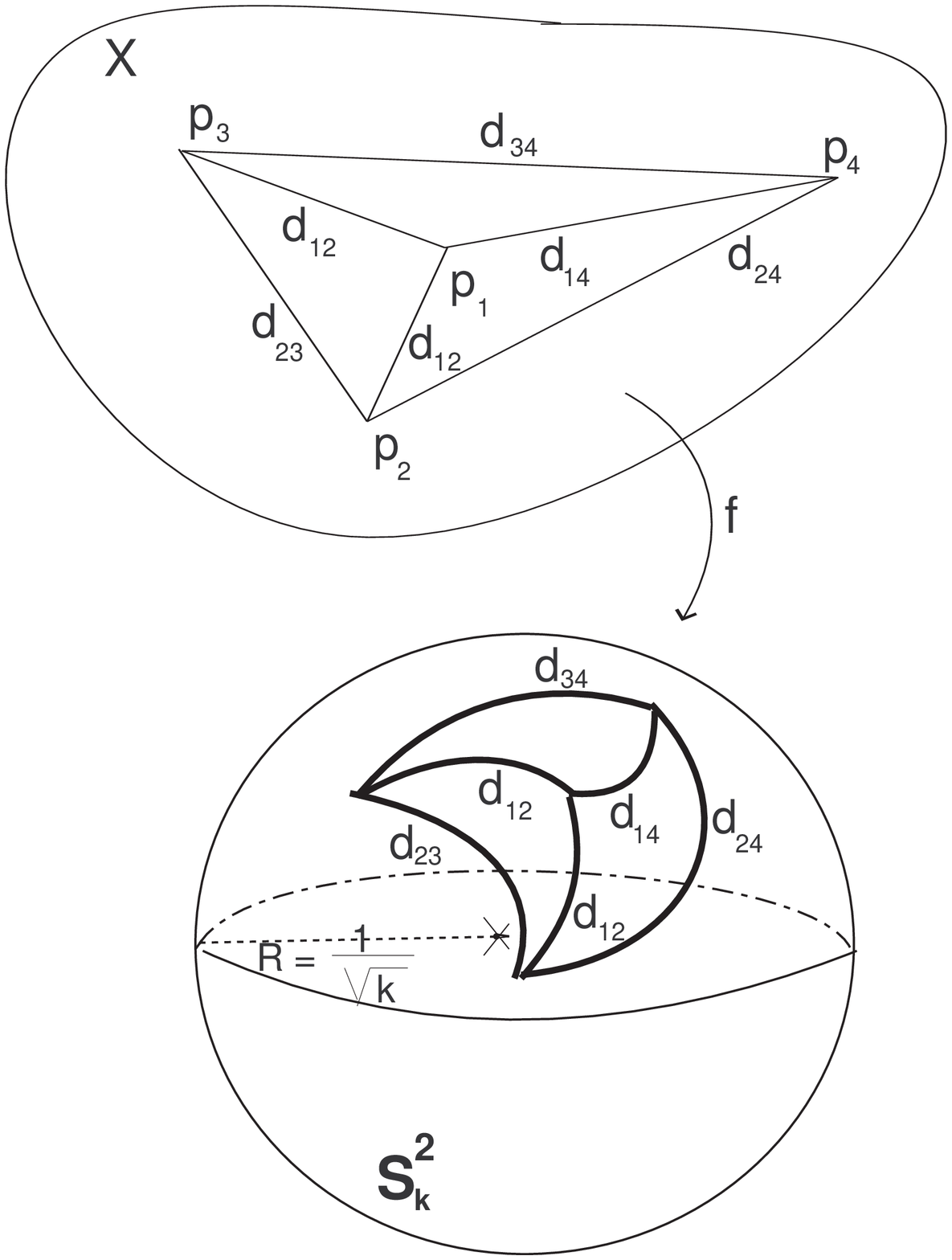}
\end{center}
\caption{}
\end{figure}

\begin{figure}[b]
\begin{center}
\includegraphics[scale=0.3]{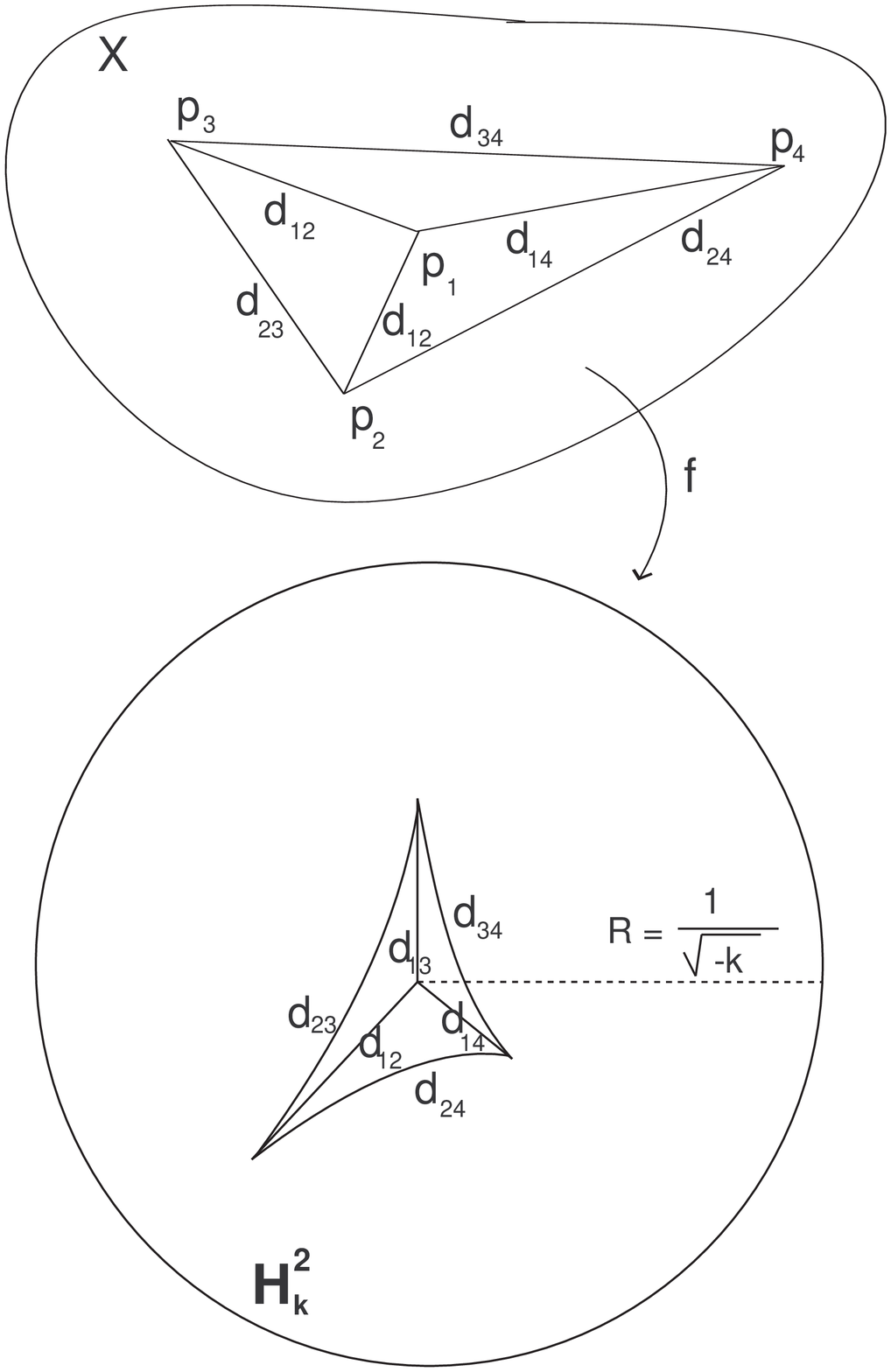}
\end{center}
\caption{}
\end{figure}

We can now define the embedding curvature at a point in a natural way by passing to the limit (but without
neglecting the existence conditions), more precisely:

\begin{defn} Let $(M,d)$ be a metric space, and let $p \in M$ be an accumulation point. Then $p$ is said to have {\em Wald
curvature} $\kappa_W(p)$ iff
\\ (i) $\exists \hspace{-0.2cm}/\, \,N  \in \mathcal{N}(p), N $ linear\footnote{\,The neighborhood $N$ of $p$ is called linear iff $N$ is contained in a geodesic.}\,;
\\ (ii) $\forall \, \varepsilon > 0, \; \exists \, \delta > 0$ s.t. $Q = \{p_1,...,p_4\} \subset M\,,$ and s.t. $d(p,p_i) < \delta \,(i=1,...,4)$
\\ $\Longrightarrow |\kappa(Q) - \kappa_W(p)| < \varepsilon$.
\end{defn}

\begin{rem}
\begin{enumerate}
\item If one uses the second (abstract) definition of the metric curvature of quadruples, then the very existence
of $\kappa(Q)$ is not assured, as it is shown by the following
\begin{cntxmp} The metric quadruple of lengths  \[d_{12} = d_{13} = d_{14} = 1; \;d_{23} = d_{24} = d_{34} = 2\]
admits no embedding curvature.
\end{cntxmp}
\item Even if a quadruple has an embedding curvature, it still may be not unique (even if $Q$ is not liniar),
indeed, one can study the following examples:
\begin{exmp}
\begin{enumerate}
\item
The quadruple $Q$ of distances $d_{ij} = \pi/2, 1 \leq i < j \leq 4$ is isometrically embeddable both in $S_0 =
\mathbb{R}^2$ and in $S_1 = \mathbb{S}^2$.
\item The quadruple $Q$ of distances $d_{13} = d_{14} = d_{23} = d_{24} = \pi,\; d_{12} = d_{34} = 3\pi/2$ admits
exactly two embedding curvatures: $\kappa_1 \in (1.5,2)$ and $\kappa_2 = 3$. (See \cite{bm}.)
\end{enumerate}
\end{exmp}
\end{enumerate}
\end{rem}

However, for "good" metric spaces\footnote{\,i.e. spaces that are locally "plane like"} the embedding curvature
exists and it is unique. And, what is even more relevant for us, this embedding curvature coincides with the
classical Gaussian curvature. The proof of this result is rather long and tedious, therefore we shall present here
only  a brief sketch of it. (This will prove to be somewhat redundant anyhow, in view of the more general results
presented in the previous section, a fact but we shall emphasize later in our presentation).)
\\The Main ingredient for this proof, and for the analysis of yet another another approach to curvature (the CAT
one) is provided by the following string of propositions (which are just generalizations of the well known
high-school triangle inequalities):
\begin{prop} Let $\triangle(p_1,q_1,r_1) \subset \mathcal{S}_{\kappa_1}$ and $\triangle(p_2,q_2,r_2) \subset \mathcal{S}_{\kappa_2}$, s.t.
\\ $p_1q_1 = p_2q_2, \; p_1r_1 = p_2r_2$ and $\angle(q_1,p_1,r_1) = \angle(q_2,p_2,r_2)$.
\\ Then:  $\kappa_1 < \kappa_2 \Longrightarrow q_1r_1 > q_2r_2$.
\end{prop}

\begin{prop} Let $p_1,q_1,r_1 \in \mathcal{S}_{\kappa_1}, \; p_2,q_2,r_2 \in \mathcal{S}_{\kappa_2}$ two isometric triples of points, s.t. the triple
$p_1,q_1,r_1$is not linear. Then:
\\ $\angle(q_1,p_1,r_1) < \angle(q_2,p_2,r_2), \angle(p_1,q_1,r_1) < \angle(p_2,q_2,r_2), \angle(q_1,r_1,p_1) <
\angle(q_2,r_2,p_2)$.
\end{prop}

\begin{prop} Let $Q_1 = \{p_1,q_1,r_1,s_1\},\, Q_2 = \{p_2,q_2,r_2,s_2\}$ be  non-linear and non-degenerate quadruples in $\mathcal{S}_{\kappa_1},\, \mathcal{S}_{\kappa_2}$, respectively.
If $\triangle(p_1,q_1,r_1)  \cong \triangle(p_2,q_2,r_2)$ and $\kappa_1 < \kappa_2$, then:
\begin{enumerate}
\item $p_1s_1 = p_2s_2,\; q_1s_1 = q_2s_2 \Longrightarrow r_1s_1 > r_2s_2$\,;
\item $r_1s_1 = r_2s_2,\; q_1s_1 = q_2s_2 \Longrightarrow p_1s_1 > p_2s_2$\,;
\item $p_1s_1 = p_2s_2,\; r_1s_1 = r_2s_2 \Longrightarrow q_1s_1 < q_2s_2$\,.
\end{enumerate}
\end{prop}

In order that we fully exploit the results above we need the following definition:
\begin{defn}
A metric quadruple $Q = Q(p_1,p_2,p_3,p_4)$, of distances \\$d_{ij} = dist(p_i,p_j), \; i=1,...,4$, is called {\em
semi-dependent} (a {\em sd-quad}, for brevity), iff 3 of its points are on a common geodesic, i.e. there exist 3
indices -- e.g. 1,2,3 -- s.t.: $d_{12} + d_{23} = d_{13}$.
\end{defn}

Now we can easily formulate the following immediate consequence of Prop. 3.10.\,:
\begin{cor} A sd-quad admits at most one embedding curvature.
\end{cor}

Unfortunately -- as we have already noticed -- in the general case the uniqueness of the embedding curvature is
not guaranteed. However we can be a bit more explicit using the following definition:
\begin{defn} Let $Q =  \{p,q,r,s\}$ be a non-linear and non-degenerate quadruple.
Q is called {\em planar} iff $\angle(q,p,r) +  \angle(q,p,s) +  \angle(s,p,r) = 2\pi$.
\end{defn}
Then we have
\begin{prop} Let $Q =  \{p,q,r,s\}$ be a a non-linear and non-degenerate quadruple in $\mathcal{S}_{\kappa}$. Then
\begin{enumerate}
\item If $Q$ is planar, then it admits no isometric embedding in $\mathcal{S}_{\kappa_1}, \; \kappa_1 > \kappa$.
\item If $Q$ is not planar, then it admits no isometric embedding in $\mathcal{S}_{\kappa_2}, \; \kappa_2 < \kappa$.
\end{enumerate}
\end{prop}

\begin{cor} Let $Q =  \{p,q,r,s\}$ be a a non-linear and non-degenerate quadruple. Then $Q$ has at most two
different embedding curvatures.
\end{cor}

In fact we can state a much stronger assertion, of which Example 3.7.(a) is just a very particular case:

\begin{prop} $\forall\, p \; \in \mathcal{S}_{\kappa}$, and $\forall\, \kappa > 0,\; \exists\, U \in
\mathcal{N}(p)$ s.t. $\exists$ a nonlinear, non-degenerate quadruple $Q \subset U$ of embedding curvature 0.
\end{prop}

\begin{proof} Let $\gamma_1, \gamma_2 \in U$,   two great-circle arcs s.t. $\gamma_1  \cap \gamma_2 = p$.
\\ Let $q_1,q_2 \in \gamma_1$ s.t. $pq_1 = pq_2 \neq 0$ and let $q \in \gamma_2$ s.t. $pq <
\pi/2\sqrt{\kappa}$. \footnote{\,i.e a quarter of the length of a great circle in $\mathcal{S}_{\kappa}$}
\\ Consider $\triangle(q'_1q'_2,q') \subset \mathbb{R}^2, \; \triangle(q'_1q'_2,q') \cong \triangle(q_1q_2,q)$,
let $p' = \frac{1}{2}q'_1q'_2$, and let \\ $h = q'p'$.
\begin{figure}[h]
\begin{center}
\includegraphics[scale=0.25]{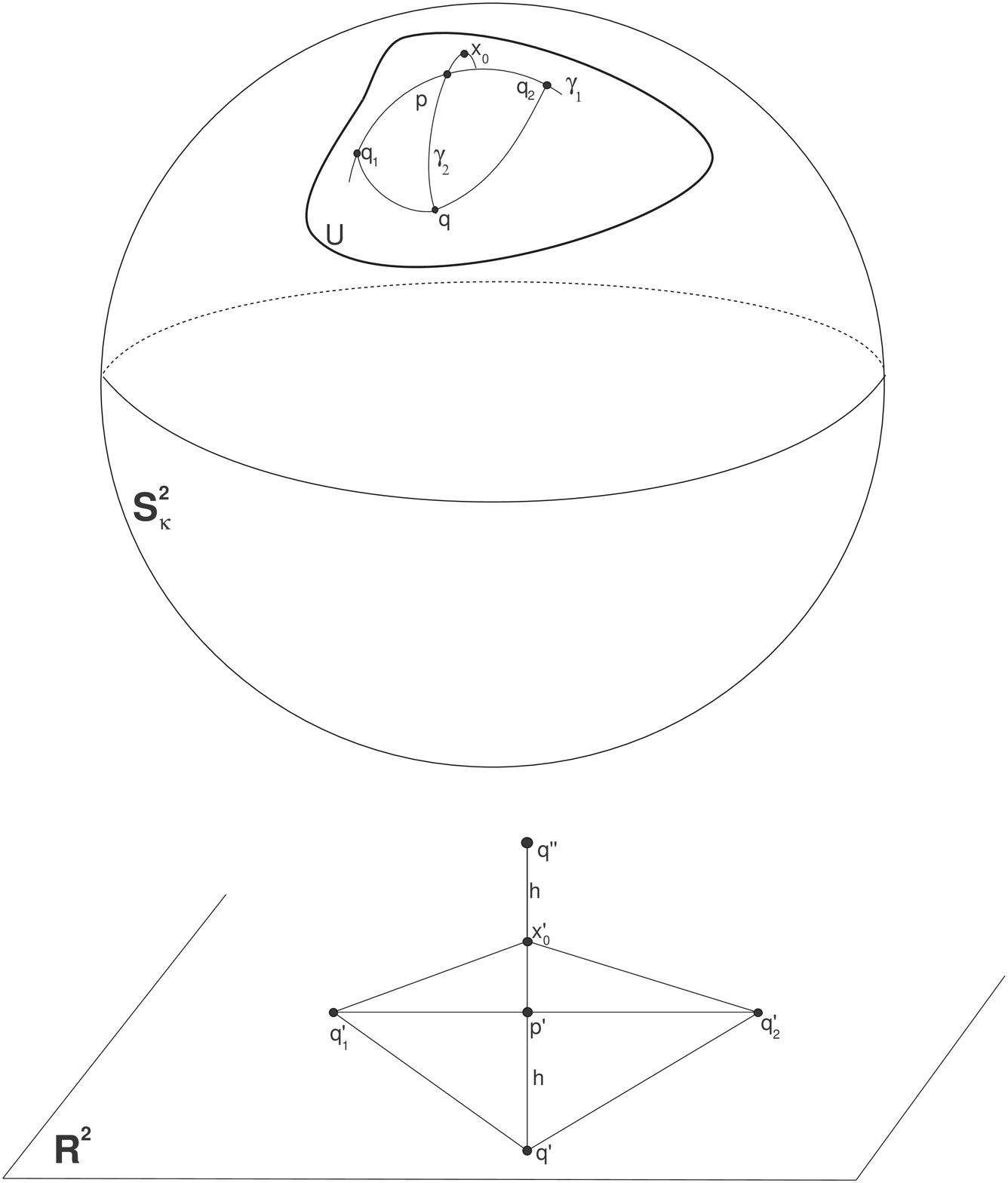}
\end{center}
\caption{}
\end{figure}

Then since $0 < \kappa$, Proposition 3.10.(3) applied to the quadruples $\{q,q_1,q_2,p\}$ and
$\{q',q'_1,q'_2,p'\}$ implies that $h < pq$.
\\ Now let $x \in \gamma_2$, $x$ between $p$ and $q$, and let $x' \in \mathbb{R}^2$ s.t. $\triangle(q'_1q'_2,x') \cong
\triangle(q_1q_2,x)$ s.t. $x$ and $q'$ are on different sides of the line
$\stackrel{\longleftrightarrow}{q_1q_2}$. Then,  $x = p \; \Rightarrow \; xq > x'q'$, and $x = q \; \Rightarrow xq
= 0 < x'q' = 2h$, where, in this case $x' = q''$. (See Figure 9.) Then it follows from a continuity argument that
$\exists\, x_0 \in \gamma_2,\; x_0$ between $p$ and $q$, s.t. $x_0q = x_0'q'$, thus implying that $\{q_1,q_2,q,x\}
\cong \{q'_1,q'_2,q',x'\}$.
\end{proof}

\begin{rem} $\{q_1,q_2,q,x\}$ is planar, while $\{q'_1,q'_2,q',x'\}$ is not planar.
\end{rem}

\subsection{The Wald Curvature vs. Gauss Curvature }
The discussion above would be nothing more than a nice intellectual exercise where it not for the fact that the
metric (Wald) and the classical (Gauss) curvatures coincide whenever the second notion makes sense, that is  for
smooth (i.e. of class $\geq \mathcal{C}^2$) surfaces in $\mathbb{R}^3$. More precisely the following theorem
holds:

\begin{thm}{\em (Wald)} Let $S \subset \mathbb{R}^3, \; S \in \mathcal{C}^m, \, m \geq 2$ be a smooth surface.
\\ Then $\kappa_W(p)$ exists, for all $p \in S$, and $\kappa_W(p) = \kappa_G(p), \forall\, p \in  M$.
\end{thm}

Moreover, Wald also proved that a partial reciprocal theorem holds, more precisely he proved the following:

\begin{thm} Let $M$ be a compact and convex metric space.
\\ If $\kappa_W(p)$ exists, for all $p \in M$, then  M is a smooth surface and $\kappa_W(p) = \kappa_G(p), \forall\, p \in \nolinebreak[4] M$.
\end{thm}

\begin{rem} I one tries to restrict oneself, in the building of Definition 3.4. only to sd-quads, then  Theorem
3.19. holds only if the following presumption is added:
\begin{cond} $M$ is locally homeomorphic to
$\mathbb{R}^2$.
\end{cond}
\end{rem}
However the proof of this facts is involved and, as such, beyond the scope of this presentation. Therefore we
shall restrict ourselves to a succinct description of the principal steps towards the proofs.
\\ The basic idea is to show that if a metric $M$ space admits a Wald curvature at any point, than $M$ is
locally homeomorphic to $\mathbb{R}^2$, thus any metric proprieties of $\mathbb{R}^2$ can be translated to $M$,
(in particular the first fundamental form).
\\ The first of these partial results is:
\begin{thm} Let $M$ be a convex metric space. Then $M$ admits at most one Wald curvature $\kappa_W(p), \; \forall \, p \in
M$.
\end{thm}

\begin{prf} By Corollary 3.12. it suffices to prove that any disk neighborhood $B(p\,;\rho) \in \mathcal{N}(p)$ contains a
non degenerate sd-quad.
\\ Without loss of generality one can assume that $B(p\,;\rho)$ contains three points $p_1,p_2,p_3$ s.t. $d(p,p_i) < \rho/2,\, i =
1,2,3$.\footnote{\,See \cite{b}} Then, by the convexity of $M$ it follows that $\exists \, q \in M$ s.t. $p \neq
p_2,p_3$ and $p_2q + p_3q = p_2p_3$. But $p_2p_3 \leq pp_2 + pp_3 < \rho \Longrightarrow (pq < \rho/2) \vee (pp_2
< \rho/2)$. In the first inequality holds, then $pq \leq pp_2 + p_2q < \rho$, i.e $q \in B(p\,;\rho)$; and if the
second one holds, then $pd \leq pp_3 + p_3q < \rho$, i.e. $q \in B(p\,;\rho)$. But $p \neq q$, therefore
$p,p_2,p_3,q$ are not linear.
\end{prf}

Our next step will be to analyze those neighborhoods that display "a normal behavior", both metrically and
curvature-wise: that is precisely those disk neighborhoods in which the Wald curvature is defined and ranges over
a small, bounded set of values prescribed by the very radius of the disk:
\begin{defn} A disk neighborhood $B(p\,;\rho); \rho > 0$ is called {\em regular} iff $\forall$  non-degenerate quadruple $Q \subset
B(p\,;\rho)$\,, $\kappa_W(Q)$ exists and $|\kappa_W(Q)| < \pi^2/16\rho^2$.
\end{defn}
\begin{rem} If $\kappa_W(p)$ exists, then for any sufficiently small $\rho$, $B(p\,;\rho)$ will be regular.
\end{rem}

It turns out that regular neighborhoods, in {\em compact, convex} spaces have the following "nice" (i.e. Real
Plane like) proprieties:
\begin{prop} Let $M$ be a compact, convex metric space and let $B(p\,;\rho) \subset M$ be a regular neighborhood.
Then if a non-degenerate quadruple $Q \subset B(p\,;\rho)$ contains two linear triples of points, then $Q$ is
linear.
\end{prop}
\begin{prop} Let $M$ be a compact, convex metric space. Then:
\\ $\forall\, p \in M$ and $\forall\, B(p\,;\rho)$ regular, $\exists \,q,r \in B(p\,;\rho)$ s.t. $p,q,r$ are not linear.
\end{prop}
\begin{prop} Any regular neighborhood $B(p\,;\rho)$ of a compact, convex metric space is {\em strictly convex}, i.e. $q,r \in B(p\,;\rho) \Longrightarrow int(qr) \subset
B(p\,;\rho)$.
\end{prop}
While the proof of this last Proposition is lengthy, that of the following important Corollary is not:
\begin{cor} Let $B(p\,;\rho)$ be a regular neighborhood. Then, $\forall\, q,r \in B(p\,;\rho),\; \exists!\, qr$ and $int(qr) \subset
B(p\,;\rho)$.
\end{cor}
\begin{proof} By the convexity of $B(p\,;\rho)$ it follows the existence of at least one geodesic $qr,\; \forall \, q,r \in
B(p\,;\rho)$. If $s \in int(q)r$, then by the proposition above we have that $s \in B(p\,;\rho)$. It follows that
$B(p\,;\rho)$ contains all the geodesics with end points $q,r$. Hence, by Proposition 3.25.\,, the geodesic
segment $qr$ is unique.
\end{proof}

We can now begin to prove that a compact, convex metric space locally mimics $\mathbb{R}^2$. We start by showing
that the {\em sinus} function is defined on $M$, thus allowing for angle measure (hence for the definition of
Polar Coordinates on regular neighbourhoods\footnote{\,In the same way geodesic polar coordinates are used on
classical surfaces.}).
\\ First, let $M$ be as before, and let $p \in M$ s.t. $\kappa_W{p}$ exists. Let $q,r \in B(p\,;\rho), q \neq p \neq r$,
where $B(p\,;\rho)$ is a regular neighborhood of $p$. Then, $\forall\, x \in [0,min\{pq,pr\})$, define $q(x) \in
pq, r(x) \in pr$ by: $d(p,q(x)) = x = d(p,r(x))$, and let $d(x) = d(,q(x),r(x))$ (see Figure 10 bellow).
\begin{figure}[h]
\begin{center}
\includegraphics[scale=0.25]{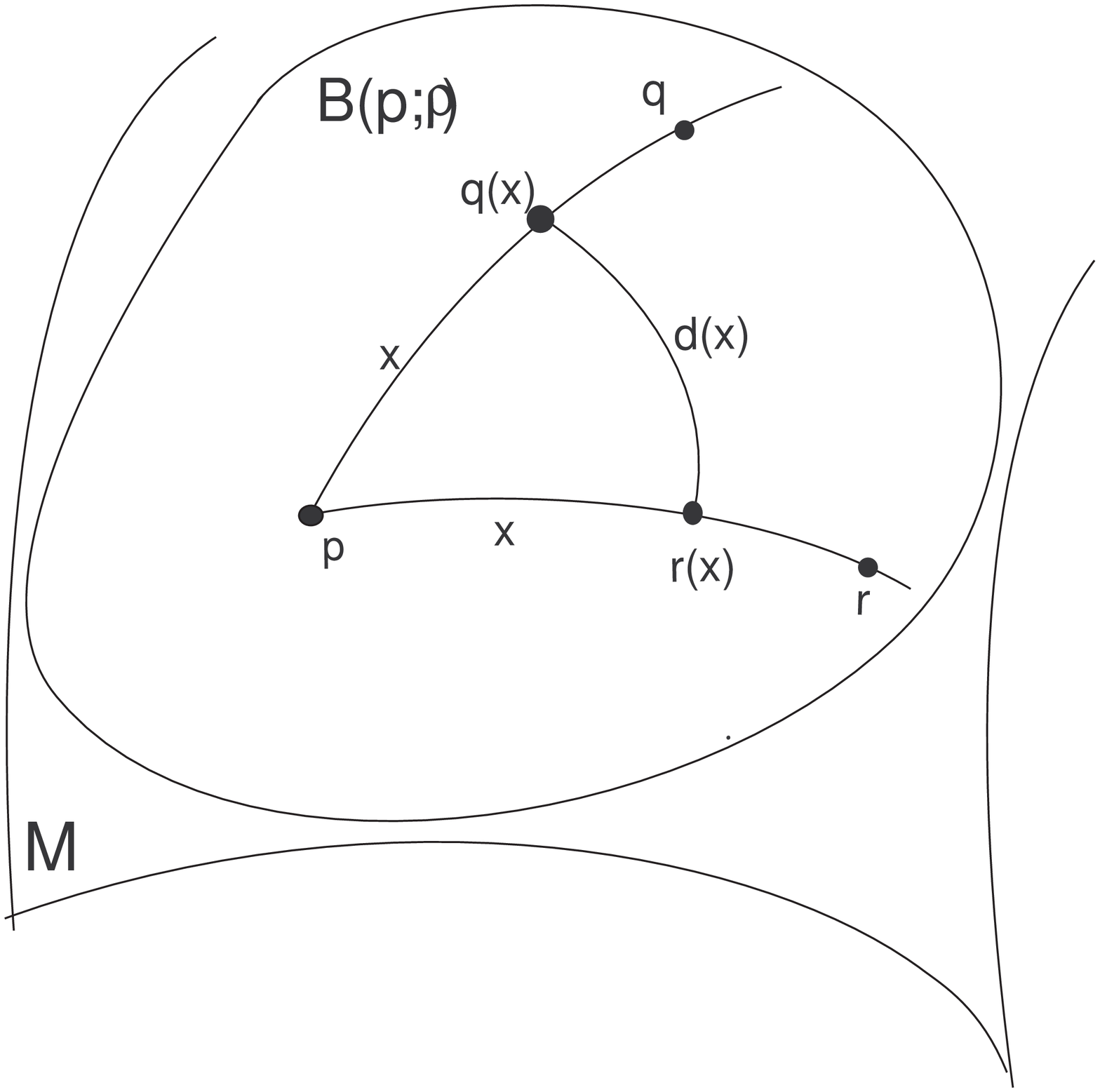}
\end{center}
\caption{}
\end{figure}
\begin{prop} The following limit exists:\[\lim_{x \rightarrow 0}{\frac{d(x)}{x}}\,.\]
\end{prop}
We omit the proof since it is rather involved (but canonical for any axiomatic approach to Euclidian Geometry --
see, for instance, \cite{b}, \cite{rr}.)

Now we can define the measure of angles at $p$\,:
\begin{defn} The measure of the angle $\angle(q,p,r))$ is given by:
\[m(\angle(q,p,r)) \eqdef 2arcsin\Big(\frac{1}{2}\lim_{x \rightarrow 0}{\frac{d(x)}{x}}\Big)\,.\]
\end{defn}

\begin{rem} The Definition above enables us to define Polar Coordinates on regular neighborhoods\footnote{\,... and once Coordinates (be they Polar or Cartesian) are introduced, the (local) homomorphism with $\mathbb{R}^2$ is immediate.}
in the following manner:
\\ Let $p_1,p_2 \in B(p\,;\rho)$ s.t. $p,p_1,p_2$ are not collinear. (Such points exist by Proposition \nolinebreak[4]3.26.). To every point $q \in
B(p\,;\rho)$ we associate the following pair of real numbers (defining the Polar Coordinates of $q$ relative to
the frame determined by $p,p_1,p_2$): $(r(q),\theta(q))$, where \[r(q) \eqdef d(p,q)\] and
 \[\theta(q)) \eqdef \left\{
         \begin{array}{clclcr}
           \mbox{$m(\angle(q,p,p_1))$} &  \mbox{if $|m(\angle(p_2,p,p_1)) - m(\angle(q,p,p_1))| = m(\angle(q,p,p_1))$ \,;} \\
           \mbox{$2\pi - m(\angle(q,p,p_1))$} &  \mbox{if $|m(\angle(p_2,p,p_1)) - m(\angle(q,p,p_1))| \neq m(\angle(q,p,p_1))$ \,.}
           \end{array}
           \right .  \]
\end{rem}

We can now safely state the foretold homomorphism result:
\begin{prop} Any convex, compact metric space is locally homeomorphic to the real plane.
\end{prop}

\section{Computing Embedding Curvature} In this section we develop formulas for the computation of Embedding
Curvature of Quadruples. First we follow the classical approach of Wald-Blumenthal that employs the so-called {\em
Cayley-Menger determinants} (see bellow). Unfortunately, the formulas obtained, albeit precise are transcendental.
Therefore we present, in the next subsection, the approximate formulas developed by C.V. Robinson.
\subsection{Embedding Curvature -- The Determinant Approach} Given a general metric quadruple $Q =
Q(p_1,p_2,p_3,p_4)$, of distances \\$d_{ij} = dist(p_i,p_j), \; i=1,...,4$, we denote by  $D(Q) =
D(p_1,p_2,p_3,p_4)$ the following determinant:
\begin{equation}
 D(p_1,p_2,p_3,p_4) = \left| \begin{array}{ccccc}
                                            0 & 1 & 1 & 1 & 1 \\
                                            1 & 0 & d_{12}^{2} & d_{13}^{2} & d_{14}^{2} \\
                                            1 & d_{12}^{2} & 0 & d_{23}^{2} & d_{24}^{2} \\
                                            1 & d_{13}^{2} & d_{23}^{2} & 0 & d_{34}^{2} \\
                                            1 & d_{14}^{2} & d_{24}^{2} & d_{34}^{2} & 0
                                      \end{array}
                               \right|
\end{equation}
Then the {\it embedding curvature} $\kappa(Q)$ of $Q$ is given -- depending upon the embedding space (i.e. upon
the sign of the curvature) -- by the following formulae:

\begin{equation}
 \kappa(Q) = \left\{
         \begin{array}{clclcrcr}
           \mbox{0} &  \mbox{if $D(Q) = 0$\,;} \\
           \mbox{$\kappa,\, \kappa < 0$} & \mbox{if $det({\cosh{\sqrt{-\kappa}\cdot d_{ij}}}) = 0$\,;} \\
           \mbox{$\kappa,\, \kappa > 0$} & \mbox{if $det(\cos{\sqrt{\kappa}\cdot d_{ij}})$ and $\sqrt{\kappa}\cdot d_{ij} \leq
           \pi$}\\
           & \mbox{\;\; and all the principal minors of order $3$ are $\geq 0$.}
         \end{array}
   \right.
\end{equation}

The determinant $D(Q) = D(p_1,p_2,p_3,p_4)$ is called the {\em Cayley-Menger determinant} (of the points
$p_1,...p_4$)\footnote{\,This definition readily generalizes to any dimension, as do the results bellow.} and, in
order to prove (4.2) we need first to investigate some of its properties.
\\ We start with the following
\begin{lem} Let: $p_1,...,p_4$ be points in $\mathbb{R}^3$. Then:
\begin{equation}
 Gram(\overrightarrow{p_1p_2},\overrightarrow{p_1p_3},\overrightarrow{p_1p_4}) =
\frac{1}{8}D(p_1,p_2,p_3,p_4)\,;
\end{equation} where
\[ Gram(\overrightarrow{p_1p_2},\overrightarrow{p_1p_3},\overrightarrow{p_1p_4}) \stackrel{def}{=}  det(\overrightarrow{p_1p_i}\cdot \overrightarrow{p_1p_j})_{i,j=2,3,4}\]
\\(Here "$\cdot$" denotes the standard scalar (dot) product in $\mathbb{R}^3$.)
\end{lem}
\begin{proof} Use expansion and manipulation of determinants.
\end{proof}
Since is a known fact that:
\[Gram(\overrightarrow{p_1p_2},\overrightarrow{p_1p_3},\overrightarrow{p_1p_4}) = 
\big(Vol(p_1,p_2,p_3,p_4)\big)^2\,;\]
where $Vol(p_1,p_2,p_3,p_4)$ denotes the (un-oriented) volume of the parallelepiped determined by the vertices
$p_1,...,p_4$ (and with edges $\overrightarrow{p_1p_2},\overrightarrow{p_1p_3},\overrightarrow{p_1p_4}$); formula
(2.3) shows that:
\begin{equation}
 D(p_1,p_2,p_3,p_4) = 8\big(Vol(p_1,p_2,p_3,p_4)\big)^2\,.
\end{equation}
Therefore the following assertion is immediate:
\begin{prop}
The points $p_1,...,p_4$ are the vertices of a simplex in  $\mathbb{R}^3$ iff $D(p_1,p_2,p_3,p_4) \neq 0$\..
\end{prop}
However, we can prove the much strong result bellow:
\begin{thm}
Let $d_{ij} > 0\,, 1\leq 4\,, i\neq j$.
\\ Then there exists a simplex $T = T(p_1,...,p_4) \subseteq \mathbb{R}^3$
s.t. $dist(x_i,x_j) = d_{ij}\,, i \neq j$; iff $D(p_i,p_j) < 0\,,(\forall)\, \{i,j\} \subset \{1,...,4\}$ and
\\$D(p_i,p_j,p_k) > 0\,, (\forall)\, \{i,j,k\} \subset \{1,...,4\}$;
\\  where, for instance,
\[D(p_1,p_2) = \left| \begin{array}{ccc}
                                            0 & 1 & 1  \\
                                            1 & 0 & d_{12}^{2} \\
                                            1 & d_{12}^{2} & 0
                                  \end{array}
                           \right| \,\]
and
 \[D(p_1,p_2,p_3) = \left| \begin{array}{cccc}
                                            0 & 1 & 1 & 1 \\
                                            1 & 0 & d_{12}^{2} & d_{13}^{2}  \\
                                            1 & d_{12}^{2} & 0 & d_{23}^{2}  \\
                                            1 & d_{13}^{2} & d_{23}^{2} & 0
                                  \end{array}
                           \right| \,;\]
etc...
\end{thm}

 In fact, the necessary and sufficient condition above can be relaxed, indeed  one can also show that the following holds\footnote{\,For the direct proof of Theorem  4.3., see \cite{be} or, alternatively  \cite{b}}:
\begin{prop}\hspace{-0.2cm}\footnote{\,We formulate this result -- for convenience and practicality -- for the case $n = 3$, only. However it is readily generalized to any dimension.}
Let $d_{ij} > 0\,, 1\leq 4\,, i\neq j$.
\\ Then there exists a simplex $T = T(p_1,...,p_4) \subseteq \mathbb{R}^3$
s.t. $dist(x_i,x_j) = d_{ij}\,, i \neq j$; iff $D(p_1,p_2,p_3,p_4) \neq 0$ and $sign\,D(p_1,p_2,p_3,p_4) = + 1$\,.
\end{prop}
\begin{proof}\hspace{-0.2cm}{\em (Sketch)} Sufficient to show (by using standard operations on determinants) that:
\[D(p_1,p_2,p_3)D(p_1,...,\hat{p_i},...,p_4) = M_{i4}^2 + D(p_1,...,\hat{p_i},...,p_4)D(p_1,p_2,p_3,p_4)\,;\]
where $M_{i4}$ is the cofactor (in $D$) of $d_{i4}^2$, and were we used the notation:
\\$\{p_1,...,\hat{p_i},...,p_4\} = \{p_1,p_2,p_3,p_4\} \setminus \{p_i\}$.
\end{proof}
Proving the formula for the spherical and hyperbolical cases would prove to be to technical for this limited
exposition; suffice to say that they essentially reproduce the proof given in the Euclidian case, and tacking into
account the fact that performing computations in the spherical (resp. hyperbolic) metric one has to replace the
distances $d_{ij}$ by $\cos{d_{ij}}$ (resp. $\cosh{d_{ij}}$)\footnote{\, See \cite{b} for the full details}.

\subsection{Embedding Curvature -- Approximate Formulas}

 The formulas we just developed in  are not only transcendental, but also the computed curvature
may fail to be unique (see preceding section). However, uniqueness is guaranteed for sd-quads. Moreover, the
relatively simple geometric setting of sd-quads allows for the development of simple  (i.e. rational) formulas for
the approximation of the Embedding Curvature.

\begin{prop} Given the metric quadruple $Q = Q(p_1,p_2,p_3,p_4)$, of distances \\$d_{ij} = dist(p_i,p_j), \;
i=1,...,4$,  the embedding curvature $\kappa(Q)$ is well approximated by:
\begin{equation}
K(Q) = \frac{6(\cos{\angle_{0}2} + \cos{\angle_{0}2'})}{d_{24}(d_{12}\sin^2({\angle_{0}2}) +
d_{23}\sin^2({\angle_{0}2'}))}
\end{equation}
 where: \(\angle_{0}2 =
\angle(p_1p_2p_4)\,, \; \angle_{0}2' = \angle(p_3p_2p_4)\) represent the angles of the  Euclidian triangles of
sides $d_{12}, d_{14}, d_{24}$ and $d_{23}, d_{24}, d_{34}$\,, respectively.
\\  The {\it error} $R$ can be estimated by using the following inequality:
\begin{equation}
|R| = |R(Q)| = |\kappa(Q) - K(Q)| < 4\kappa^{2}(Q)diam^{2}(Q)/\lambda(Q)
\end{equation}
 where we put: $ \lambda(Q) =
d_{24}(d_{12}\sin{\angle_{0}2} + d_{23}\sin{\angle_{0}2'})/S^2 $, and where $S = Max\{p,p'\}; \; \\ 2p = d_{12} +
d_{14} + d_{24}\,, \; 2p' = d_{32} + d_{34} + d_{24}$.
\end{prop}
\begin{prf} The basic idea of the proof is to recreate, in a general metric setting, the Gauss Map -- in this case
one measures the curvature by the amount of "bending" one has to apply to a general planar quadruple so that it
may be "straightened" (i.e. isometrically embedded as a {\em sd-quad}) in some $S_\kappa$.
\\ Consider two plane\footnote{i.e. embedded in $R^2 \equiv \mathcal{S}_0$} triangles $\triangle p_1p_2p_4$
and $\triangle p_2p_3p_4$, and denote by $\triangle p_{1,k}p_{2,k}p_{4,k}$ and $\triangle p_{2,k}p_{3,k}p_{4,k}$
their respective isometric embeddings into $\mathcal{S}_k$. Then $p_{i,k}p_{j,k}$ will denote the geodesic (of
$\mathcal{S}_k$) through $p_{i,k}$ and $p_{j,k}$. Also, let $\angle_k2$ and $\angle_k2'$ denote, respectively, the
following angles of $\triangle p_{1,k}p_{2,k}p_{4,k}$ and $\triangle p_{2,k}p_{3,k}p_{4,k}$\,: $\angle_k2 = \angle
p_{1,k}p_{2,k}p_{4,k}$ and $\angle_k2' = \angle p_{2,k}p_{3,k}p_{4,k}$. (See Fig.\,11)
\begin{figure}[h]
\begin{center}
\includegraphics[scale=0.3]{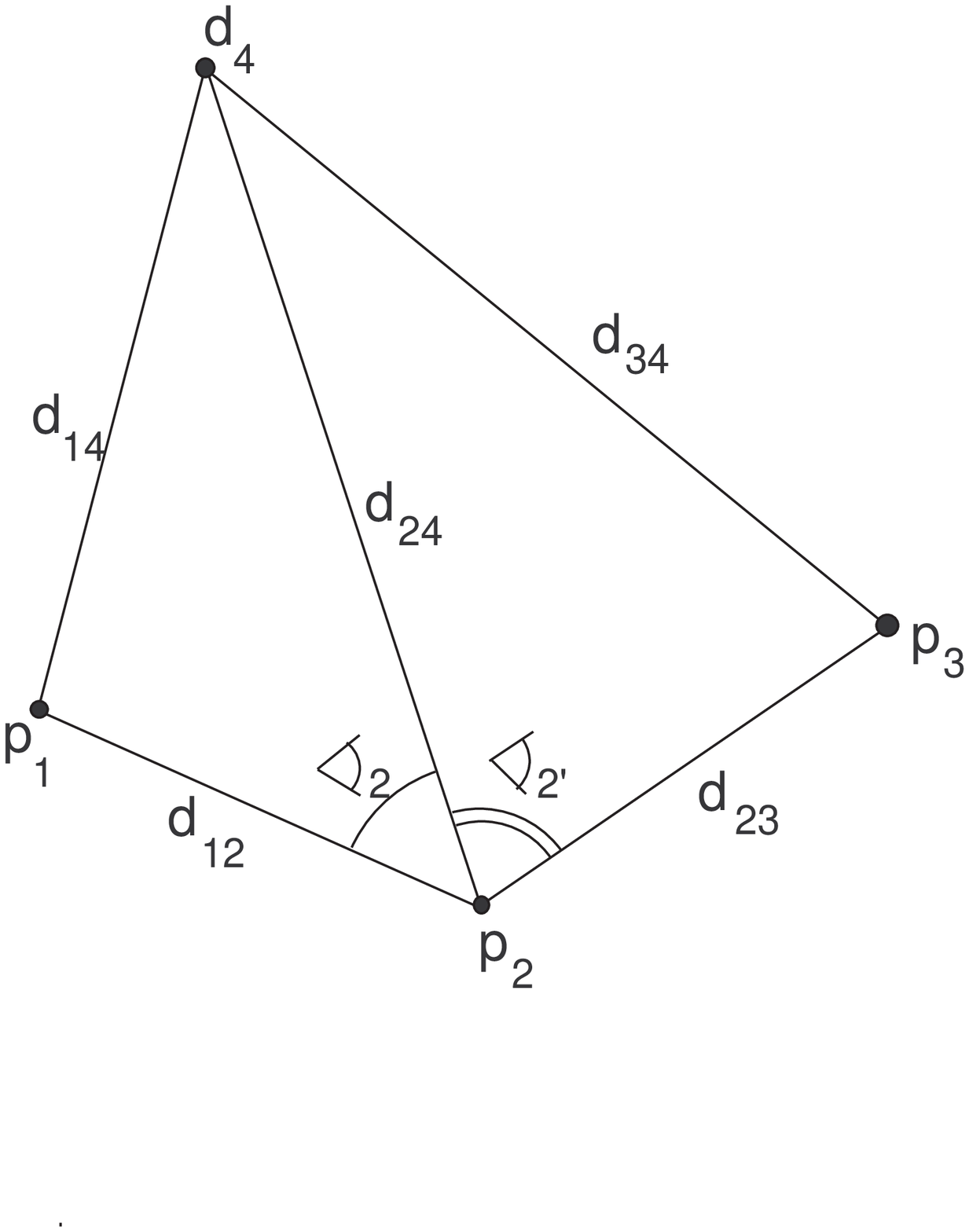}
\caption{ }
\end{center}
\end{figure}
\\ But $\angle_k2$ and $\angle_k2'$ are strictly increasing as functions of $k$. Therefore the equation
\begin{equation}
\angle_k2 + \angle_k2' = \pi
\end{equation}
has at most one solution $k^*$, i.e. $k^*$ represents the unique value for which the points $p_1,p_2,p_4$ are on a
geodesic in $\mathcal{S}_k$ (for instance on $p_1p_4$).
\\ But that means that $k^*$ is precisely the Embedding Curvature, i.e. $k^* = \kappa(Q)$\,, where $Q = Q(p_1,p_2,p_3,p_4)$.
\\  Equation $(4.7)$ is equivalent to
\[\cos^2\frac{\angle_{k^*}2}{2} \; + \; \cos^2\frac{\angle_{k^*}2'}{2}\, = \, 1\]
The basic idea being the comparison between metric triangles with equal sides, embedded in $\mathcal{S}_0$ and
$\mathcal{S}_k$, respectively, it is natural to consider instead of the previous equation, the following:
\begin{equation}
\theta(k,2)\cdot\cos^2\frac{\angle_{0}2}{2} \; + \; \theta(k,2')\cdot\cos^2\frac{\angle_{0}2'}{2}\, = \, 1
\end{equation}
 where we denote:
\[\theta(k,2) := \frac{\cos^2\frac{\angle_{k^*}2}{2}}{\cos^2\frac{\angle_{0}2}{2}}\,; \;\; \theta(k,2') := \frac{\cos^2\frac{\angle_{k^*}2'}{2}}{\cos^2\frac{\angle_{0}2}{2}}\,.\]
Since we want to approximate $\kappa(Q)$ by $K(Q)$ we shall resort -- naturally -- to expansion into MacLaurin
series. We are able to do this because of the existence of the following classical formulas:
\[ \cos^2\frac{\angle_{k}2}{2} = \frac{\sin({p\sqrt{k}})\cdot\sin({d\sqrt{k}})}{\sin({d_{12}\sqrt{k}})\cdot\sin({d_{24}\sqrt{k})}}\,; \;k > 0\,;\]
\[ \cos^2\frac{\angle_{k}2}{2} = \frac{\sinh({p\sqrt{k}})\cdot\sinh({d\sqrt{k}})}{\sinh({d_{12}\sqrt{k})}\cdot\sinh({d_{24}\sqrt{k}})}\,; \;k <0\,; \]
and, of course
\[\cos^2\frac{\angle_{0}2}{2} = \frac{pd}{d_{12}d_{24}}\,;\]
were:  $d = p - d_{14} =  (d_{12} + d_{24} - d_{14})/2\,.$\footnote{and the analogous formulas for
$\cos^2\frac{\angle_{k'}2}{2}$.}
\\ By using the development into series of $f_1(x) = \frac{\sin{\sqrt{x}}}{{\sqrt{x}}}$ and $f_2(x) =
\frac{\sinh{\sqrt{x}}}{{\sqrt{x}}}$; one (easily) gets the  desired expansion for $\theta(k,2)$:
\begin{equation}
\theta(k,2) = 1 + \frac{1}{6}kd_{12}d_{24}\big(\cos(\angle_0{2}) - 1\big) + r\,;
\end{equation}
where: $|r| < \frac{3}{8}k^2p^4$\,, for $|kp^{2}| < 1/16$\,. 
 By applying (4.9.) to (4.8), we receive:
\begin{equation}
[1 + \frac{1}{6}k^*d_{12}d_{24}\big(\cos(\angle_0{2}) - 1\big) + r]\cos^2\frac{\angle_{0}2}{2} \;+\;
\end{equation}
\[[1 +  \frac{1}{6}k^*d_{23}d_{24}\big(\cos(\angle_0{2'}) - 1\big) + r']\cos^2\frac{\angle_{0}2'}{2} = 1\,;\]
for: $|r| + |r'| < \frac{3}{4}(k^*)^2(Max\{p,p'\})^4 = \frac{3}{4}(k^*)^2S^4$\,.
\\ By solving linear equation (in variable $k^*$) (4.10) and using some elementary trigonometric transformation one has:
\[k^* = \frac{6(\cos{\angle_0{2}} + \cos{\angle_0{2'}})}{d_{24}(d_{12}\sin^2({\angle_0{2}}) + d_{23}\sin^2({\angle_0{2'}}))} + R\]
where:
\[|R| < \frac{12(|r| +  |r'|)}{d_{24}\big(d_{12}\sin^2({\angle_0{2}}) + d_{23}\sin^2({\angle_{0}2'})\big)} < \frac{9(k^*)^2\max\{p,p'\}}{d_{24}\big(d_{12}\sin^2({\angle_0{2}}) + d_{23}\sin^2({\angle_{0}2'})\big)}\]
But $k^* \equiv \kappa(Q)$ so we get the desired formula (4.5)\,.
\\ To prove the correctness of the bound (4.6) one has only to observe that:
\[S = Max\{p,p'\} < 2diam(Q), \;\big(diam(Q) = \max_{1\leq i<j\leq4}\{d_{ij}\}\big),\]
and perform the necessary arithmetic manipulations.
\end{prf}
\begin{rem}
(a) The function $\lambda = \lambda(Q)$ is continuous and 0-homogenous as a function of the $d_{ij}$-s. Moreover:
$\lambda(Q) \geq 0$ and $\lambda(Q) = 0 \Leftrightarrow \sin{\angle_{0}2} = \sin{\angle_{0}2'} = 0$, i.e. iff $Q$
is linear. [Therefore for sd-quads $\lambda(Q) > 0$ and, moreover, $\lambda(Q) \rightarrow 0 \Rightarrow \\ Q
\rightarrow linearity $.]
\\ \hspace*{2cm} (b) Since $\lambda(Q) \neq 0$ it follows that: $K(Q) \in \mathbb{R}$ for any quadrangle $Q$.
\\ In addition: $sign(k(Q)) = sign(K(Q))$.
\\ \hspace*{2cm} (c) If $Q$ is any sd-quad, then $\kappa^{2}(Q)diam^{2}(Q)/\lambda(Q) < \infty$. Moreover, if $\lambda(Q) \gg \hspace{-0.4cm}{/} \;\:0$,\footnote{i.e. for not very small values of $\lambda(Q)$} then $\kappa^{2}(Q)diam^{2}(Q)/\lambda(Q)\gg
\hspace{-0.4cm}{/} \;\:0\,$ i.e. $|R|$ is {\em small} if $Q$ is not close to linearity.
\\ In this case $|R(Q)| \sim diam^{2}(Q)$ (for any given $Q$).
\end{rem}
Since the Gaussian curvature $k_G(p)$ at a point $p$ is given by: \[k_G(p) = \lim_{n \rightarrow 0}{\kappa(Q_n)}
\,;\] where $Q_n \rightarrow Q = \Box p_1pp_3p_4\,; \; diam(Q_n) \rightarrow 0$,
 from {\em Remark 4.6.}(c) we  immediately infer that the following holds\footnote{ and gives theoretical
justification to the algorithm}:
\begin{thm}
Let $S$ be a differentiable surface. Then, for any point $p \in S$:
\[k_G(p) = \lim_{n \rightarrow 0}{K(Q_n)}\, ;\]
for any sequence $\{Q_n\}$ of sd-quads that satisfy the following condition: \[Q_n \rightarrow Q = \Box
p_1pp_3p_4\,; \; diam(Q_n) \rightarrow 0 \,.\]
\end{thm}

\begin{rem}
 In the following special cases even "nicer" formulas are obtained:
\begin{enumerate}
\item  If $d_{12} = d_{32}$, then
\begin{equation}
K(Q) = \frac{12}{d_{13}\cdot d_{24}}\cdot\frac{\cos{\angle_{0}2} + \cos{\angle_{0}2'}}{\sin^2{\angle_{0}2} +
\sin^2{\angle_{0}2'}}\,;
\end{equation}
(here we have of course: $d_{13} = 2d_{12} = 2d_{32}$); or, expressed as a function of distances alone:
\begin{equation}
K(Q) = 12\frac{2d_{12}^2+2d_{24}^2-d_{14}^2-d_{13}^2}{8d_{12}^2d_{24}^2 - (d_{12}^2 +d_{24}^2-d_{14}^2)^2 -
(d_{12}^2 +d_{24}^2-d_{34}^2)^2}
\end{equation}
\item If $d_{12} = d_{32} = d_{24}$ and if the following condition also holds:
\item $\angle_{0}2' = \pi/2$; i.e. if $d_{34}^2 = d_{12}^2 + d_{24}^2$ or, considering (2), also: $d_{34}^2 = 2d_{12}^2$
\\ then
\begin{equation}
K(Q) = \frac{6\cos{\angle_{0}2}}{d_{12}(1+\sin^2{\angle_{0}2})} =
\frac{2d_{12}^2-d_{14}^2}{4d_{12}^4+4d_{14}^2d_{12}^2-d_{14}^4}\,.
\end{equation}
\end{enumerate}
\end{rem}

\section{Appeendix 1 -- The Menger and Haantjes Curvatures}
Better known than the Wald Curvature, the Menger Curvature is a metric definition of curvature of {\em curves}, as
is the Haantjes Curvature. As such they can be employed as {\em sectional curvatures} to approximate curvature of
triangulated surfaces. \\ We begin by introducing the the Menger Curvature: this is a metric expression for the
circum-radius of a triangle\footnote{\,thus giving in the limit a metric definition of the Osculatory Circle},
based upon elementary high-school formulas:
\begin{defn} Let $(M,d)$ be a metric space, and let $p,q,r \in M$ be three distinct points. Then:
\[K_M(p,q,r) = \frac{\sqrt{(pq+qr+rp)(pq+qr-rp)(pq-qr+rp)(-pq+qr+rp)}}{pq \cdot qr \cdot rp}\,; \]
is called the {\em Menger Curvature} of the points $p,q,r$.
\end{defn}

We can now define the Menger Curvature at a given point by passing to the limit:
\begin{defn} Let (M,d) be a metric space and let $p \in M$ be an accumulation point. Then $M$ has at $p$ {\em Menger
Curvature} $\kappa_M(p)$ iff
\\  $\forall \, \varepsilon > 0, \; \exists \, \delta > 0$ s.t. $d(p,p_i) < \delta\,;\; i=1,2,3\;\Longrightarrow |K(Q) - \kappa_M(p)| < \varepsilon$.
\end{defn}

\begin{rem} The apparent equivalent notion of {\em Alt} Curvature, in which one uses only two points converging to the third, is in fact more general, where we  define the
Arp curvature by:
\begin{defn} Let (M,d) be a metric space and let $P \in M$ be an accumulation point. Then $M$ has at $p$ {\em Alt
Curvature} $\kappa_A(p)$ iff the following limit exists
\[\kappa_A(p) \eqdef \lim_{q,r \rightarrow p}K(p,q,r)\,.\]
\end{defn}
\end{rem}

However, both $\kappa_M(p)$ and $\kappa_A(p)$ suffer from the same imperfection: since they are both modelled
closely after the Euclidian Plane, they convey this Euclidian type of curvature upon the space they are defined
on. However, the next definition doesn't mimic closely $\mathbb{R}^2$ so it better fitted for generalizations:

\begin{defn} Let (M,d) be a metric space and let $c: I=[0,1] \stackrel{\sim}{\rightarrow} M$ be a homeomorphism, and let $p,q,r \in c(I),\; q,r \neq p$. Denote by
$\widehat{qr}$ the arc of $c(I)$ between $q$ and $r$, and by $qr$ segment from $q$ to $r$. (See Figure 12 bellow.)
\begin{figure}[h]
\begin{center}
\includegraphics[scale=0.25]{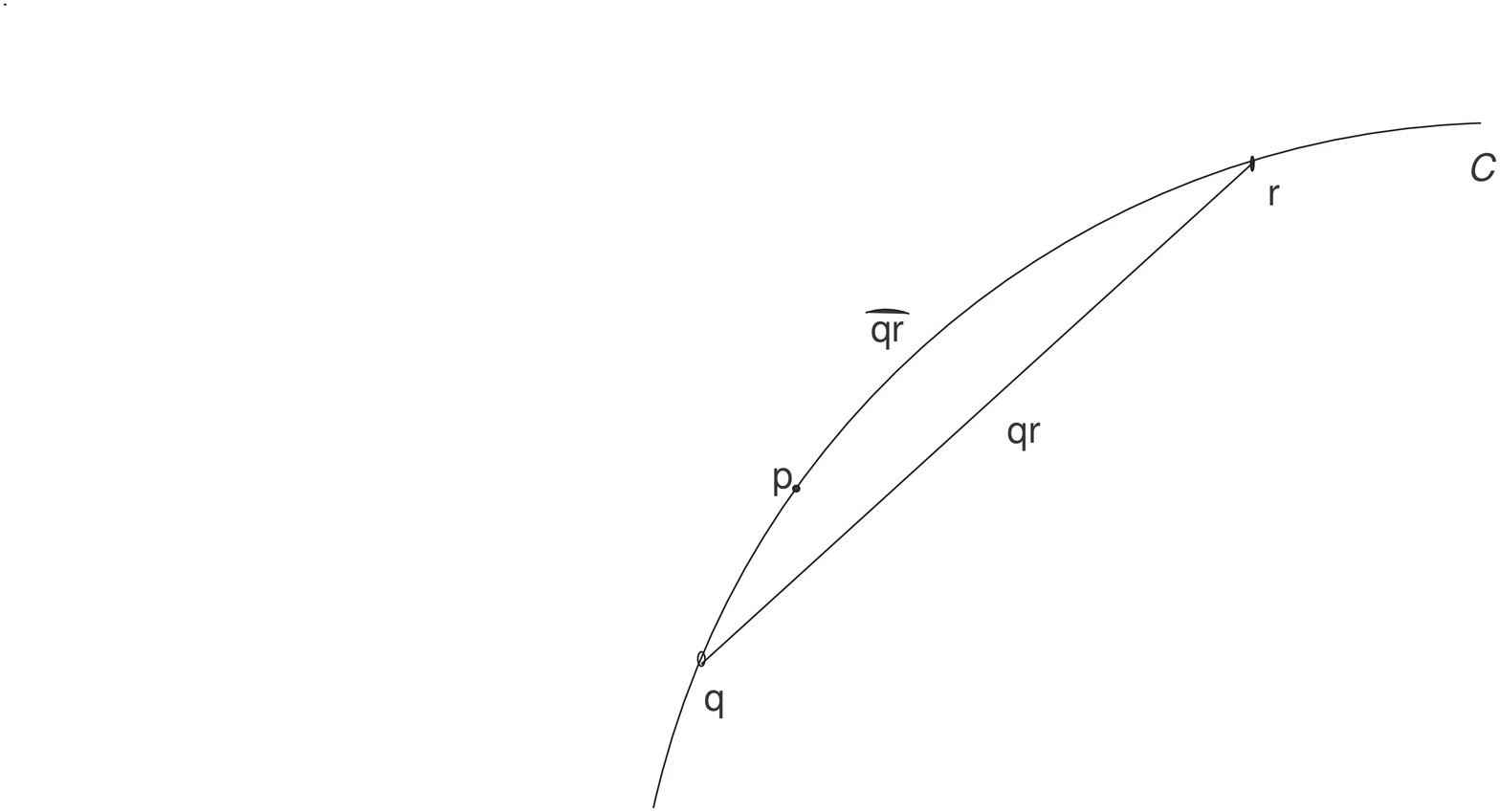}
\end{center}
\caption{ }
\end{figure}
\\ Then $c$ has {\em Haantjes Curvature} $\kappa_H(p)$ at the point $p$ iff:
\begin{displaymath}
\kappa_H^2(p) = 24\lim_{q,r \rightarrow p}\frac{l(\widehat{qr})-d(q,r)}{\big(l(\widehat{qr})\big)^3}\,\,;
\end{displaymath}
where "$l(\widehat{qr})$" denotes the length\footnote{\,given by the intrinsic metric induced by $d$} of
$\widehat{qr}$.
\end{defn}

\begin{rem} $\kappa_H$ exists only for rectifiable curves, but if $\kappa_M$  exists at any point $p$ of $c$, then
$c$ is rectifiable.
\end{rem}

\begin{rem} Evidently we have the following relationship between curvatures:
\[\exists\,\kappa_M \; \Longrightarrow \; \exists\,\kappa_A\,.\]
while
\[\exists\,\kappa_A \; \Longrightarrow\hspace{-0.5cm}/ \;\;\; \exists\,\kappa_M\,.\]
However, we can prove the following theorem:
\end{rem}

\begin{thm} Let $c:I \rightarrow M$ be a rectifiable curve, and let $p \in M$.
\\ If $\kappa_A$ (or $\kappa_M$) exists, then $\kappa_H(p)$ exists and
\[\kappa_A = \kappa_H(p)\,.\]
\end{thm}

\begin{rem} This last result and the Remark preceding it allow as to employ any of the curvatures above in
estimating curvatures of smooth curves on triangulated surfaces.
\end{rem}

\section{Appendix 2 -- The Rinow Curvature}
The curvatures introduced before may seem a bit archaic in comparison to the more fashionable approach of {\em
comparison triangles}, with their ar reaching applications. We present here one of these comparison criteria and
show its equivalence with the Wald curvature. We start with the following definition:

\begin{defn} Let $(M,d)$ be a metric space, together with the intrinsic metric induced by $d$. Let $R = int(R) \subseteq M$ be a region
of $M$. We say that $R$ is a {\em region of  curvature} $\leq \kappa$ ($\kappa \in \mathrm{R}$) iff
\begin{enumerate}
\item $\forall p,q \in R\;, \exists$ a geodesic segment $pq \subset R$;
\item $\forall \, T(p,q,r) \subset R$ is isometrically embeddable  in $\mathcal{S}_{\kappa}$;
\item If $T(p,q,r) \subset R$ and $x \in pq, y \in pr$, and if the points $p_\kappa,q_\kappa,r_\kappa,x_\kappa,y_\kappa \in \mathcal{S}_{\kappa}$ satisfy the following conditions:
\begin{enumerate}
\item $T(p,q,r) \cong T(p_\kappa,q_\kappa,r_\kappa)$;
\item $T(p,q,x) \cong T(p_\kappa,q_\kappa,x_\kappa)$;
\item $T(p,r,y) \cong T(p_\kappa,r_\kappa,y_\kappa)$;
\end{enumerate}
then $xy \leq x_{\kappa}y_\kappa$.
\end{enumerate}
By replacing the condition: "$xy \leq x_{\kappa}y_\kappa$" with: "$xy \geq x_{\kappa}y_\kappa$", we obtain the
definition of a {\em region of  curvature} $\geq \kappa$. (See Fig.\,13.)
\end{defn}

\begin{figure}[h]
\begin{center}
\includegraphics[scale=0.25]{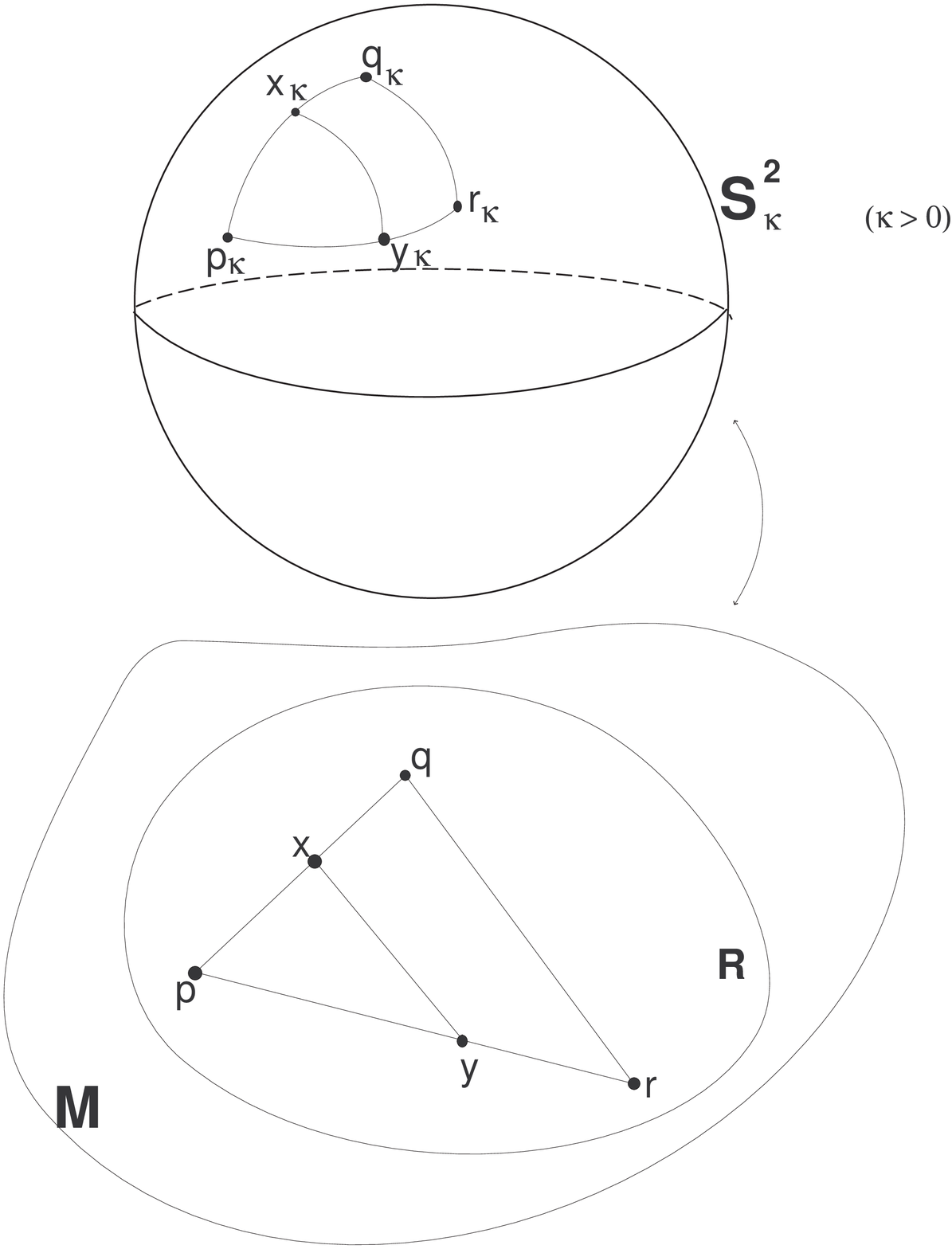}
\end{center}
\caption{ }
\end{figure}

We now pass to the localization of the Definition above:
\begin{defn} Let $(M,d)$ be a metric space, together with the intrinsic metric induced by $d$, and let $p \in M$
be an accumulation point. Then $M$ has at $p$ {\em Rinow Curvature} $\kappa_R(p)$ iff
\\ (i) $\exists \hspace{-0.2cm}/\, \,N  \in \mathcal{N}(p),\, N $ linear;
\\ (ii) $\forall\, \varepsilon > 0,\; \exists\, \delta . 0$, s.t.$ B(p;\delta)$ is  (a) a region of Rinow curvature $\leq \kappa_R(p) + \varepsilon$ and
(b)  a region of Rinow curvature $\geq \kappa_R(p) - \varepsilon$.
\end{defn}

While its greater generality endows the Rinow curvature with more flexibility in applications and makes it easier
in generalization, it is even more difficult to compute than Wald Curvature. However this quandary was has an
almost ideal solution, due to Kirk (see \cite{k}), solution which we briefly expose here:

\begin{defn} Let $M$ be a compact, convex metric space, and let $p \in M$.
\\ If $\kappa_W(p)$ exists, then $\kappa_R(p)$ exists, and $\kappa_R(p) = \kappa_W(p)$.
\end{defn}

Unfortunately, since $\kappa_R(p)$ makes no presumption of dimensionality, the existence of $\kappa_R(p)$ does not
imply the existence of $\kappa_W(p)$.
\begin{cntxmp} Let $M \equiv \mathbb{R}^3$. Then $\kappa_R(p) \equiv 0$ but $\kappa_W(p)$ does not exist at any
point, since every neighborhood contains linear quadruples.
\end{cntxmp}

The solution (due to Kirk) of this problem is to consider the {\em Modified Wald curvature} $\kappa_{WK}$, defined
as follows:
\begin{defn} Let $(M,d)$ be a metric space, together with the intrinsic metric induced by $d$, and let $p \in M$.
Then $M$ has at $p$ {\em Modified Wald curvature Curvature} $\kappa_{WK}(p)$ iff
\\ (i)  $\exists \hspace{-0.2cm}/\, \,N  \in \mathcal{N}(p),\, N $ linear;
\\ (ii) $\forall\, \varepsilon > 0,\; \exists\, \delta . 0$, s.t. if $Q \subset B(p;\delta)$ is a non-degenerate
sd-quad, then $\kappa_W(Q)$  exists and $|\kappa_{WK}(p)  - \kappa_W(Q)| < \varepsilon$.
\end{defn}

\begin{rem} $\exists\,\kappa_W(p) \; \Longrightarrow \; \exists\,\kappa_{WK}(p)$ but $\exists \kappa_{KW}(p) \Longrightarrow\hspace{-0.5cm}/\;\;\; \kappa_{W}(p)$.
\end{rem}

This modified curvature indeed represents the wished for solution, as proved by the following to Theorems:
\begin{thm} Let $(M,d)$ be a metric space. Then:
\\ \hspace*{2cm} $\exists\,\kappa_R(p) \; \Longrightarrow \; \exists\,\kappa_{WK}(p)$ and $\kappa_R(p) =
\kappa_{WK}(p)$.
\end{thm}

\begin{thm} Let $(M,d)$ be a metric space together with the associated intrinsic metric, and let $p \in M$. Then,
if
\\ (i) $\kappa_{WK}(p)$;
\\ and if
\\ (ii) $\exists \, B(p;\rho) \in \mathcal{N}(p)$, s.t. $qr \subset B(p;\rho), \; \forall q,r \in B(p;\rho)$;
\\ then $\kappa_R(p)$ exists and $\kappa_R(p) = \kappa_{WK}(p)$.
\end{thm}

\section{Appendix 3 -- The Radius Formula}
The Cayley-Menger determinant allows one to express not only the volume and area\footnote{ The 2-dimensional
analogue of Formula (4.4) for the area of the triangle $T(p_1,p_2,p_3)$ being: \[\big(Area\,(p_1,p_2,p_3)\big)^2 =
-D(p_1,p_2,p_3)\,.\] } of simplices in $\mathbb{R}^n$ but (as expected) it may be used to compute the radius of
the circumscribed sphere around an Euclidian simplex. To be more precise, we have the following result\footnote{
that can be readily generalized to higher dimensions}:
\begin{thm}
\begin{enumerate}
\item
The radius $R = R(p_1,p_2,p_3,p_4)$ of the sphere circumscribed around the tetrahedron $T(p_1,p_2,p_3,p_4) \in \mathbb{R}^3$ is given by: 
\[R^2 = -\frac{1}{2}\frac{\Delta(p_1,p_2,p_3,p_4)}{D(p_1,p_2,p_3,p_4)} \]
where:
\[\Delta(p_1,p_2,p_3,p_4) =  \left| \begin{array}{cccc}
                                                 0 & d_{12}^{2} & d_{13}^{2} & d_{14}^{2} \\
                                                 d_{12}^{2} & 0 & d_{23}^{2} & d_{24}^{2} \\
                                                 d_{13}^{2} & d_{23}^{2} & 0 & d_{34}^{2} \\
                                                 d_{14}^{2} & d_{24}^{2} & d_{34}^{2} & 0
                                    \end{array}
                               \right| \]

\item The points $p_1,p_2,p_3,p_4,p_5 \in \mathbb{R}^3$ are coplanar or co-spherical iff
\[\Delta(p_1,p_2,p_3,p_4,p_5) = 0\;.\]
\end{enumerate}
\end{thm}

\begin{prf}
\begin{enumerate}
\item If $p_0 \in\mathbb{R}^3$ is s.t. $d_{0i} = R,\; i=1,...,5$, then by direct computation we obtain:
\[\Gamma(p_0,...,p_5) = -2R^2\Gamma(p_1,...,p_5) - \Delta(p_1,p_2,p_3,p_4,p_5)\,;\]
from which the desired formula follows immediately if we chose $p_0$ as the center of the sphere circumscribed
around the points $p_1,p_2,p_3,p_4,p_5$.
\item
\end{enumerate}\vspace*{-1.5em} \hspace*{1.3cm} $(\Longrightarrow)$ Let $\{x^1,x^2,x^3\}$ be any orthonormal coordinate frame for $\mathbb{R}^3$, and let $p_i^j;
\; i=1,...,5; j=1,2,3$; represent the coordinates of the points $p_1,p_2,p_3,p_4,p_5$ relative to this coordinate
system. Then $p_1,p_2,p_3,p_4,p_5$ belong to the same sphere or plane iff $\exists\; (a,b,c_j) \neq (0,0,0),\;
j=1,2,3$; s.t.
\[a||p_i||^2 + b + \sum_{j=1}^{3}c_jp_i^j\,; \; i=1,...,5.\]
Then $\Delta_1 = \Delta_2 = 0$, where:
\[\Delta_1(p_1,p_2,p_3,p_4,p_5) = \left| \begin{array}{ccccc}
                                                 ||p_1||^2 & 1 & p_1^1 &  p_1^2 & p_1^3\\
                                                 ||p_2||^2 & 1 & p_2^1 &  p_2^2 & p_2^3 \\
                                                 ||p_3||^2 & 1 & p_3^1 &  p_3^2 & p_3^3 \\
                                                 ||p_4||^2 & 1 & p_4^1 &  p_4^2 & p_4^3 \\
                                                 ||p_5||^2 & 1 & p_5^1 &  p_5^2 & p_5^3
                                    \end{array}
                               \right| \]
and
\[\Delta_2(p_1,p_2,p_3,p_4,p_5) =  \left| \begin{array}{ccccc}
                                                 1 & ||p_1||^2 & -2p_1^1 &  -2p_1^2 & -2p_1^3\\
                                                 1 & ||p_2||^2 & -2p_2^1 &  -2p_2^2 & -2p_2^3 \\
                                                 1 & ||p_2||^2 & -2p_3^1 &  -2p_3^2 & -2p_3^3 \\
                                                 1 & ||p_2||^2 & -2p_4^1 &  -2p_4^2 & -2p_4^3 \\
                                                 1 & ||p_2||^2 & -2p_5^1 &  -2p_5^2 & -2p_5^3
                                    \end{array}
                               \right| \]
Therefore $\Delta_1 \cdot \Delta_2^t = 0$. But $\Delta_1 \cdot \Delta_2^t = \Delta(p_1,p_2,p_3,p_4,p_5)$, so this
implication is proven.
\\ \;\; $(\Longleftarrow)$ $\Delta(p_1,p_2,p_3,p_4,p_5) = 0 \; \Longrightarrow \; \Delta_1 = 0$ and there exist
numbers $(a,b,c_j) \neq (0,0,0),\; j=1,2,3$; s.t.
\[a||p_i||^2 + b + \sum_{j=1}^{3}c_jp_i^j = 0\,; \; i=1,...,5;\]
i.e. $p_1,p_2,p_3,p_4,p_5$ belong to the plane or the sphere given by the equation
 \[a||X||^2 + b + c\sum_{j=1}^{3}c_jX = 0\,;\; X = (x_1,x_2,x_3)\,.\]
\end{prf}

\end{document}